\documentclass[preprint,a4paper,12pt]{elsarticle}

\usepackage{lineno,hyperref}
\usepackage{color}
\usepackage{multirow}
\modulolinenumbers[5]

\usepackage{graphicx}
\usepackage[top=30truemm,bottom=30truemm,left=25truemm,right=25truemm]{geometry}
\usepackage{amsmath}
\usepackage{lscape}
\usepackage{amsfonts}
\usepackage{bm}
\usepackage{algorithm}
\usepackage{algorithmic}
\usepackage{here}
\usepackage[top=30truemm,bottom=30truemm,left=25truemm,right=25truemm]{geometry}
\definecolor{green}{rgb}{0.4660, 0.6740, 0.1880}

\journal{Elsevier}
\graphicspath{{./figures/}}

\begin{document}

\begin{frontmatter}

\title{Assessments of epistemic uncertainty using Gaussian stochastic weight averaging for fluid-flow regression}

\author{Masaki Morimoto}
\address{Department of Mechanical Engineering, Keio University, Yokohama, 223-8522, Japan}
\author{Kai Fukami}
\address{Department of Mechanical Engineering, Keio University, Yokohama, 223-8522, Japan}
\address{Department of Mechanical and Aerospace Engineering, University of California, Los Angeles, CA 90095, USA}
\author{Romit Maulik\corref{mycorrespondingauthor}}
\address{Mathematics and Computer Science Division, Argonne National Laboratory, Illinois 60439, USA}
\cortext[mycorrespondingauthor]{Corresponding author}
\ead{rmaulik@anl.gov}
\author{Ricardo Vinuesa}
\address{FLOW, Engineering Mechanics, KTH Royal Institute of Technology, Stockholm, SE-100 44, Sweden}
\author{Koji Fukagata}
\address{Department of Mechanical Engineering, Keio University, Yokohama, 223-8522, Japan}

\begin{abstract}
We use Gaussian stochastic weight averaging (SWAG) to assess the epistemic uncertainty associated with neural-network-based function approximation relevant to fluid flows.
SWAG approximates a posterior Gaussian distribution of each weight, given training data, and a constant learning rate.
Having access to this distribution, it is able to create multiple models with various combinations of sampled weights, which can be used to obtain ensemble predictions.
The average of such an ensemble can be regarded as the `mean estimation', whereas its standard deviation can be used to construct `confidence intervals', which enable us to perform uncertainty quantification (UQ) with regard to the training process of neural networks.
We utilize representative neural-network-based function approximation tasks for the following cases: (i) a two-dimensional circular-cylinder wake; (ii) the DayMET dataset (maximum daily temperature in North America); (iii) a three-dimensional square-cylinder wake; and (iv) urban flow, to assess the generalizability of the present idea for a wide range of complex datasets.
SWAG-based UQ can be applied regardless of the network architecture, and therefore, we demonstrate the applicability of the method for two types of neural networks: (i) global field reconstruction from sparse sensors by combining convolutional neural network (CNN) and multi-layer perceptron (MLP); and (ii) far-field state estimation from sectional data with two-dimensional CNN.
We find that SWAG can obtain physically-interpretable confidence-interval estimates from the perspective of epistemic uncertainty.
This capability supports its use for a wide range of problems in science and engineering.
\end{abstract}


\end{frontmatter}

\section{Introduction}
\label{sec:introduction}

Neural networks (NNs) have emerged as promising tools in a broad range of applications.
Among these, the potential for effective uses of NNs for nonlinear systems has been recognized by scientists including fluid dynamicists~\cite{BNK2020}.
The strong nonlinearities and chaotic nature of fluid motion make it difficult to approximate input-output maps for fluid flows and subsequently cause bottlenecks in prediction, optimization, and control. 
Therefore, neural networks, which are excellent nonlinear function approximators, have recently become a popular tool for various fluid dynamics tasks~\cite{THBSDBDY2020,Duraisamy2021,BHT2020,font2021deep}. 
Furthermore, neural networks have also proven their utility for inverse problems, which are valuable for bridging the gap between physical experiment and numerical simulation in fluid dynamics. 
This has mostly been achieved by relating the sparse information available in the former (for example sparse sensor measurements and plane information of volumetric data) to the well-resolved flow field of the latter~\cite{NFF2021,GDISAV2021}.

For the purpose of state estimation of fluid flows, Erichson et al.~\cite{erichson2020} used a multi-layer perceptron (MLP) to reconstruct flow fields around a circular cylinder, the sea surface temperature, and forced isotropic turbulence from sparse sensor measurements. 
They demonstrated the capability of NNs for estimating a flow field from limited measurements in a computationally efficient manner. 
While the investigation above focused on a situation with a fixed number and location of sensors, Fukami et al.~\cite{FukamiVoronoi} have recently proposed a Voronoi tesselation-assisted approach in reconstructing a global field from sensors, the location and number of which can vary.
By interpreting sparse sensor measurements as low-resolution images, super-resolution analysis, which estimates a high-resolution image from low-resolution data, can also be considered as a candidate for flow-field reconstruction.
The first application of super-resolution reconstruction to fluid flow data was performed by Fukami et al.~\cite{FFT2019a}.
Subsequently, it has been used for extracting high-resolution data from coarse-grained low-resolution flow fields in many subsequent studies~\cite{GDISAV2021,kim2020deep,FFT2021b,DHLK2019,GSW2021}.
For the purpose of field estimation from sectional data, Guastoni et al.~\cite{LGIDSAV2020} utilized convolutional neural networks (CNNs) \cite{LBBH1998} to estimate a turbulent channel flow from wall information.
A similar problem setting was also considered to compare the difference between NN and linear stochastic estimation by Nakamura et al.~\cite{NFF2021}.
As shown above, NN-based spatial state estimators have been applied to numerical data~\cite{SP2019,ZXCLCPV2020,KL2020,FFT2020,MFZF2020,matsuo2021supervised}, as well as to physical experiments~\cite{CHBH2006,BEF2019,HLC2019,CZXG2019,MFF2021}.

In addition to the state estimation efforts, the applications of NN-based approximators to closure modeling of Reynolds-averaged Navier--Stokes (RANS) simulation and large-eddy simulation (LES) also exemplify the opportunities of NN-based modeling for turbulent flows~\cite{DIX2019,MSJC2019,MS2017,MSRB2019,WHP2018,wang_turbulencemodeling2017}.
Ling et al.~\cite{LKT2016} introduced the so-called {\it tensor-basis neural network} for closure modeling of RANS simulations.
Their novel approach successfully embedded invariance into the model structure by projecting the Reynolds-stress tensor onto an appropriately chosen tensor basis. 
CNN-based turbulence modeling for LES was also proposed by Lapeyre et al.~\cite{lapeyre2019training} and was compared with the MLP-based methods by Pawar et al.~\cite{pawar2020priori}.

Reduced-order modeling is also a promising use of NN-based methods~\cite{omata2019,MFF2019,fukami2020sparse,FNF2020,ASBN2020,CFPN2020,ARSRN2019}.
While closure models focus on a particular subcomponent of the traditional simulation techniques, some NN studies attempt to establish a complete surrogate model for systems of equations.
For instance, a number of studies~\cite{mohan2018deep,SGASV2019,pawar2019deep,ahmed2020long,eivazi2020recurrent,CFKNP2020,HFPN2019} have utilized long short-term memory (LSTM) to predict the temporal evolution of compressed representations of the high-dimensional flow-field state.
Furthermore, Hasegawa et al.~\cite{HFMF2020a} combined a CNN-based autoencoder with an LSTM to predict future states of a flow around various bluff-body shapes by only following the dynamics of the low-dimensional latent vector. 
A similar approach was also demonstrated for the hyperbolic shallow-water equations~\cite{maulik2021reduced}, a cylinder wake at various Reynolds numbers ($Re$)~\cite{HFMF2020b} and minimal turbulent channel flow~\cite{nakamura2020extension}.

As reviewed above, there have been several fundamental studies of NNs for various fluid-mechanics applications.
For enhancing their usefulness in practical settings where training data may be limited or difficult to acquire {\it a priori}, it is important to quantify various forms of uncertainty when making predictions with them.
The uncertainty in NNs can generally be classified into two groups: (i) data (or aleatoric) uncertainty and (ii) epistemic uncertainty~\cite{der2009aleatory, senge2014reliable,kendall2017uncertainties,hullermeier2021aleatoric}.
The data uncertainty relies on the probability distribution of a dataset, which may be associated with the inherent noise in the data-generation or observation processes.
Therefore, this uncertainty is irreducible.
For fluid-dynamics applications, Maulik et al.~\cite{MFRFT2020} have recently considered the application of probabilistic neural networks (PNNs) for reduced-order modeling and state estimation of unsteady flows and geophysical data.
The PNN assumed that targets could be generated from an underlying probabilistic process that was a mixture of Gaussians and could characterize the data uncertainty.
Predictions obtained from this framework can therefore provide associated confidence intervals for regions where larger variance was observed in the training-data distributions.

The second category of uncertainty, epistemic uncertainty, is caused due to the lack of confidence in making a prediction by a specific model~\cite{der2009aleatory,hullermeier2021aleatoric}.
From the context of NNs, this type of uncertainty which relates to the shortage of knowledge~\cite{dubois1996representing} may be reduced by generating more representative training data.
One of the well-known methods to assess epistemic uncertainty is the use of Bayesian neural networks (BNNs)~\cite{bayesian2012}, which assume each trainable parameter of an NN to be sampled from a joint distribution of all parameters, that is conditioned on the training data.
However, an experimental determination of this joint distribution, generally via sampling methods such as Markov-chain Monte--Carlo methods~\cite{MCMC2003}, is prohibitively expensive.
Therefore, several approximations are constructed for BNNs such as with Monte--Carlo dropout~\cite{MCdropout2016}, probabilistic backpropagation with variational inference~\cite{PBP2015,ricardo_VAE2021}, ensemble-based methods~\cite{BART2007,WWX2016}, and physics-informed methods~\cite{xiao_UQRANS2016,WSH2016,SW2020}.
These methods accelerate the process of estimating the aforementioned joint distribution by making various simplifying assumptions (a full review of these and other associated techniques is outside the scope of this article).
By characterizing the trained weights and biases of an NN with a density, several forward model evaluations may be made for the same set of inputs that are used for generating ensemble statistics such as mean and confidence intervals for the target variable.
In this study, we utilize a state-of-the-art epistemic uncertainty-quantification techniques, Gaussian stochastic weight averaging (SWAG)~\cite{SWAGarxiv2019}, to perform NN-based regression for various high-dimensional fluid flow estimations.
SWAG approximates the posterior distribution of the collected weights as a Gaussian distribution, conditioned on the training data.
By sampling the weights from the distribution and performing an ensemble estimation, both the target estimates and their standard deviations can be obtained.

Our assessment for SWAG is performed on a diverse set of fluid-dynamics regression tasks.
We first demonstrate the capability of SWAG for global field reconstruction from limited sensors with a two-dimensional cylinder wake at various Reynolds numbers.
We then perform a similar experiment, i.e., flow reconstruction from sparse sensors, but for the DayMET dataset which provides long-term, continuous, and gridded estimates of daily weather and climatology variables by interpolating and extrapolating ground-based observations through statistical modeling techniques.
From DayMET, we focus on the reconstruction of the daily maximum temperature over North America.
We also consider the problem of far-field flow-state estimation from cross-sectional data using a three-dimensional square cylinder wake~\cite{matsuo2021supervised} and urban flow~\cite{urbanflow}.

The present paper is organized as follows.
The concept of SWAG and machine learning models used in this study are introduced in section~\ref{sec:method}.
The applications of SWAG for a wide range of fluid flows are demonstrated in section~\ref{sec:results}.
We finally provide concluding remarks with considerable outlooks in section~\ref{sec:conclusion}.


\section{Methods}
\label{sec:method}

\subsection{Gaussian stochastic weight averaging (SWAG)}
\label{sec:swa}

\begin{figure}
    \centering
    \includegraphics[width=\textwidth]{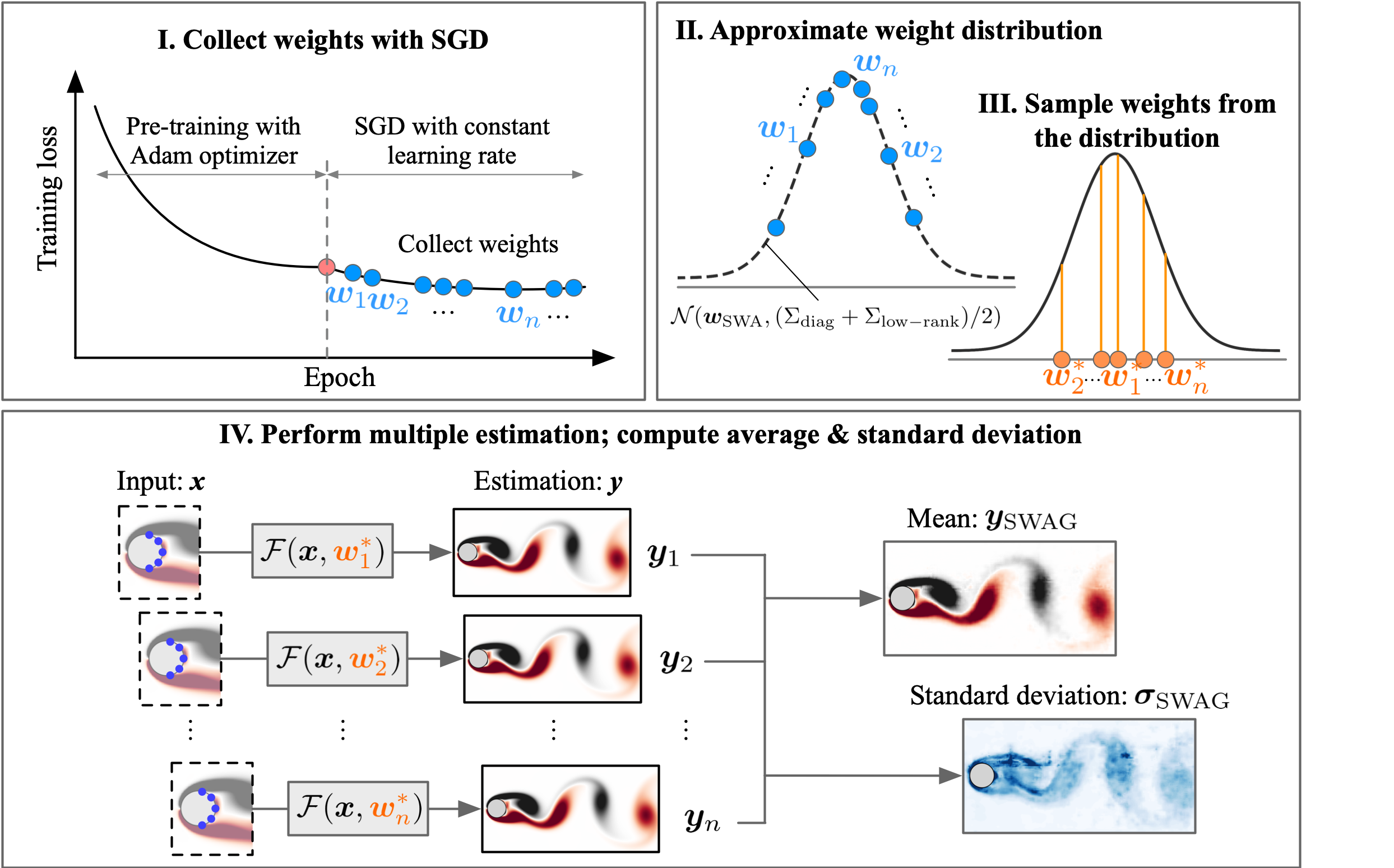}
    \caption{A high-level schematic outlining the use of SWAG-based UQ for regression using deep neural networks. }
\label{fig:concept}
\end{figure}

First, we introduce the concept of Gaussian stochastic weight averaging (SWAG)~\cite{SWAGarxiv2019} used for deep learning-based uncertainty quantification in the present study. 
SWAG is founded on the concept of stochastic weight averaging (SWA)~\cite{SWAarxiv2019}, which samples several weights during an NN training to obtain ensemble predictions. 
The detailed procedure of SWA is as follows:
Given a loss function $E$, weights inside the NN are updated via backpropagation~\cite{kingma2014} with a chosen optimizer.
Although there are various types of optimizers, most of them are based on stochastic gradient descent (SGD) to update the weights regarding a gradient of the loss function, such that, ${\bm w}_{n+1}={\bm w}_n-\eta \nabla E$, where ${\bm w}$ represents the weights, $n$ is an index of the learning epoch, and $\eta$ is a learning rate. 
In the traditional approach, the set of weights, which achieves the lowest error, is then utilized for deterministic predictions.
However, Izmailov et al.~\cite{SWAarxiv2019} claim that SGD with a high constant learning rate explores several possible high-performing solutions rather than converging to a single optimal solution.
Therefore, the primary strategy of SWA is to average the weights obtained via SGD with a constant learning rate and obtain a more generalizable solution for unseen test situations.
Note that training to convergence, usually with the Adam optimizer~\cite{kingma2014}, is performed in advance of the aforementioned weight-collection process.
SGD with a constant learning rate is then used for the further training of the deep NN, starting from the optimized solution through the pre-training.
The varying weights over this second phase of the training process are then collected to obtain an average:
\begin{equation}
    {\bm w}_{\rm SWA}=\frac{1}{N}\sum_{n=1}^{N}{\bm w}_n,
\end{equation}
where $N$ is the number of total epochs in the second training. SWA has successfully been utilized for a wide range of applications~\cite{SWAappl1,SWAappl2,SWAappl3} and has produced competitive estimates of epistemic uncertainty when compared to more traditional techniques such as Bayesian model averaging~\cite{wilson2020bayesian}.

As an extension of SWA, Gaussian stochastic weight averaging (SWAG) was introduced by Maddox et al.~\cite{SWAGarxiv2019} to obtain a covariance of the weights in addition to a mean value obtained from the ensembles.
Through this process, a Gaussian approximation could be constructed for the weights sampled near the convergence region.
The main idea of SWAG-based UQ, summarized in figure~\ref{fig:concept}, is to approximate a Gaussian distribution for each weight.
This corresponds to assuming a joint distribution of all trainable parameters with diagonal-covariance structure. 
Once this distribution is approximated, samples can be made for each forward pass given input data to obtain ensemble estimates for mean and standard deviations of the target.
Specifically, the average of the Gaussian distribution for the weights is given by ${\bm w}_{\rm SWA}$ while its variance is computed by ensembling two types of variance: $\Sigma_{\rm diag}$ and $\Sigma_{\rm low-rank}$, such that an approximated posterior distribution is represented as ${\cal N}({\bm w}_{\rm SWA},(\Sigma_{\rm diag}+\Sigma_{\rm low-rank})/2)$, as shown on the top-right panel of figure~\ref{fig:concept}.
SWAG-diagonal $\Sigma_{\rm diag}$, defined in Maddox et al.~\cite{SWAGarxiv2019}, is composed of diagonal values for the covariance matrix of the weights over all the training epochs:
\begin{equation}
    \Sigma_{\rm diag}={\rm diag}(\overline{\bm w^2}-{\bm w}_{\rm SWA}^2).
\end{equation}
In contrast, $\Sigma_{\rm low-rank}$ represents a variance of the weights during the last $L$ epochs of training:
\begin{equation}
    \Sigma_{\rm low-rank}=\frac{1}{L-1}\sum_{i=1}^L({\bm w}_{N-L+i}-{\bm w}_{\rm SWA})^2,
\end{equation}
where $N$ is the total number of epochs and $L$ is a hyperparameter for SWAG.
After obtaining a posterior distribution of the weights, we then perform a weight sampling to obtain new weights ${\bm w}^\ast$ (shown as the orange dots in figure~\ref{fig:concept}) from the distribution by taking a summation of the standard deviation of the weights and their mean value:
\begin{equation}
    {\bm w}^\ast = {\bm w}_{\rm SWA}+\frac{1}{\sqrt{2}}\Sigma_{\rm diag}^{\frac{1}{2}}r_1+\frac{1}{\sqrt{2}}\Sigma_{\rm low-rank}^{\frac{1}{2}}r_2,
\end{equation}
where $r_1$ and $r_2$ are randomly chosen from a Gaussian distribution ${\cal N}(0,1)$ for each sampling.
An average (a solution of SWAG, ${\bm y}_{\rm SWAG}$) and a standard deviation (confidence interval of the estimation, ${\bm \sigma}_{\rm SWAG}$) of the multiple estimations are then obtained as follows:
\begin{equation}
    {\bm y}_{\rm SWAG}=\frac{1}{n}\sum_{i=1}^n{\cal F}({\bm x;{\bm w}^\ast_i}),~{\bm \sigma}_{\rm SWAG}=\left[\frac{1}{n-1}\sum_{i=1}^n\left({\cal F}({\bm x};{\bm w}^\ast_i)-{\bm y}_{\rm SWAG}\right)^2\right]^{\frac{1}{2}},
    \label{eq:5}
\end{equation}
where $n$ represents the number of models sampled from a posterior distribution.

{It is also worth pointing out that the posterior $p(w | D)$ of SWAG is approximately Gaussian~\cite{mandt2017stochastic} as stochastic gradient descent (SGD) converges with fixed learning rate $\eta$. 
The posterior is given by 
\begin{align}
p(w | D) \propto \exp{\left(-\frac{N}{2} \bar{w}^T \Sigma^{-1} \bar{w}\right)}
\end{align}
where $N \in \mathbb{Z}$ is the number of batches in one training iteration and $\bar{w}, \Sigma$ are the empirical mean and covariance, respectively, of the SGD iterated updates on the vector of weights and biases of the NN $w$.
Assume that, near convergence, the gradient updates for each minibatch of SGD may be decomposed as an empirical mean and an uncorrelated Gaussian noise,
\begin{align}
\nabla_w L(w) \approx g_{\bar{w}} L + \frac{1}{\sqrt{S}} \Delta g (\bar{w}),
\end{align}
where $g_{\bar{w}} L = \nabla_{\bar{w}} L$ and $\Delta g (\bar{w}) \sim \mathcal{N} (\mathbf{0}, C)$ and $S$ is the size of the SGD minibatch. 
Since $C$ is assumed to be constant with respect to $w$, we may factorize 
\begin{align}
    C = B B^T.
\end{align}
Subsequently, near convergence, we can define 
\begin{align}
    \Delta w = w^{k+1} - w^k
\end{align}
which through the expression of SGD given by
\begin{align}
    w^{t+1} = w^t - \eta \nabla_w L(w)
\end{align}
becomes
\begin{align}
    \Delta w = -\eta g (\bar{w}) + \frac{\eta}{\sqrt{S}} B \mathcal{N} (\mathbf{0}, \mathbf{I}).
\end{align}
Assuming gradients of the loss function and the learning rate to be sufficiently small near convergence, the evolution of the SGD iterations may be explained by the following stochastic differential equation
\begin{align}
    d w_t = \eta g (\bar{w}) + \frac{\eta}{\sqrt{S}} B d \tilde{w}_t
\end{align}
where $d \tilde{w}_t$ is a standard Wiener process. 
Furthermore, if we assume that the loss surface after SGD convergence is approximated by a quadratic function
\begin{align}
    L (\bar{w} ) \sim \frac{1}{2} \bar{w}^T A \bar{w} = \tilde{L} (\bar{w}), 
\end{align}
where we have taken $\tilde{L}=0$ at $\bar{w}=0$ without loss of generality, we can represent the weight update process through SGD iteration by a special type of stochastic differential equation known as the Ornstein-Uhlenbeck process
\begin{align}
    dw_t = -\eta A w dt + \frac{\eta}{\sqrt{S}} B d \tilde{w}
\end{align}
with stationary distribution 
\begin{align}
    q(w) \sim \exp{\left(-\frac{1}{2} \bar{w}^T \Sigma^{-1} \bar{w}\right)}.
\end{align}
Note that SWA and SWAG perform a simple Monte-Carlo estimation of the stationary distribution given by
\begin{align}
    q(w) \approx \sum_k \delta(w=w_k)
\end{align}
where $k$ are the gradient updates in SGD.
}

\subsection{Machine-learning models}
\label{sec:MLmodel}

SWAG relies on estimating weight distributions for arbitrary neural-network architectures.
To demonstrate this, we consider two types of architectures: (i) MLP-CNN-based estimator~\cite{zhang2018application,bhatnagar2019prediction,MFZNF2021} and (ii) CNN~\cite{LGIDSAV2020,KL2020,FNKF2019,park_choi_2020}.
These two examples have respectively been used for a broad range of fluid-flow approximations.
We also compare these two models with a standard probabilistic neural network~\cite{MDN1994} which determines the uncertainty in the data.
In what follows, we introduce the different architectures and their approximation tasks.

\subsubsection{MLP-CNN-based estimator}

\begin{figure}
    \centering
    \includegraphics[width=\textwidth]{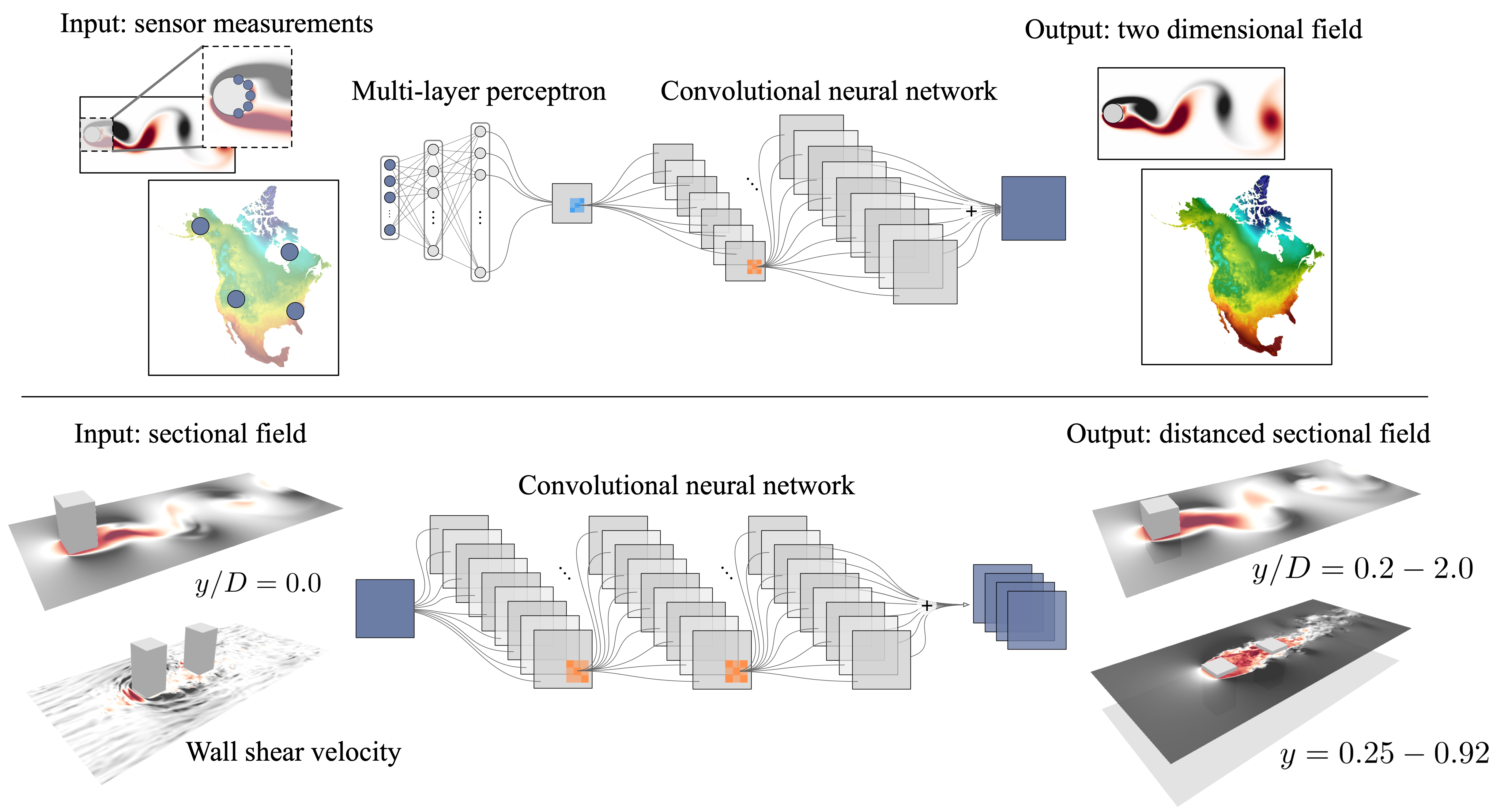}
    \caption{Machine-learning models utilized in this study. In Sec.~\ref{sec:sens2field} (above), we demonstrate SWA-based UQ with an MLP-CNN-based global-field estimator. In Sec.~\ref{sec:sec2sec} (below), a CNN is utilized for a far-field state estimation.}
\label{fig:schem}
\end{figure}

We first consider a combination of multi-layer perceptron (MLP)~\cite{RHW1986} and convolutional neural network (CNN)~\cite{LBBH1998} for a global-field-reconstruction task from sparse-sensor measurements.
For our experiments, we first use data from a cylinder wake at various Reynolds numbers and then perform similar assessments for the DayMET dataset, as illustrated in figure~\ref{fig:schem}~(top).
Local sensor measurements (and Reynolds number for a cylinder-wake example) ${\bm q}_{\rm in}$ are fed into an MLP, and then the feature vectors extracted from it are given to the convolutional layers.
This MLP-CNN-based estimator $\cal F$ is trained to optimize weights ${\bm w}$ minimizing the loss function arranged by the difference between a solution (target field) ${\bm q}_{\rm Ref}$ and an output of the model ${\cal F}({\bm q}_{\rm in})$ such that:
\begin{align}
{\bm w} = {\rm argmin}_{\bm w}||{\bm q}_{\rm Ref}-{\cal F}({\bm q}_{\rm in};{\bm w})||_2.
\end{align}
We use the $L_2$ error norm as a loss function.
As introduced in the previous section, the Adam optimizer~\cite{kingma2014} is utilized for training to attain convergence, while SGD with a constant-learning-rate schedule is used to collect weights inside the machine-learning model.

\subsubsection{Convolutional neural network}

A convolutional neural network~\cite{LBBH1998} is also constructed for the estimation of cross-sectional data from two-dimensional sectional input of flows around a square cylinder at $Re_D=300$ (which is the Reynolds number based on the free-stream velocity and the side length of the cylinder $D$) and urban-like obstacles, as illustrated in figure~\ref{fig:schem}.
Since there is no dimension reduction or expansion in this model, we use convolutional filter operations only to extract spatial features between input and output data without using pooling or upsampling operations~\cite{MFZNF2021}.
A convolutional operation can be expressed as 
\begin{equation}
    q^{(l)}_{ijg}=\varphi\left(\sum_{f=1}^F\sum_{p=0}^{\Theta-1}\sum_{q=0}^{\Theta-1}\theta^{(l)}_{pqfg}q^{(l-1)}_{i+p-C,j+q-C,f}+b_g^{(l)}\right),
    \label{eq:CNN}
\end{equation}
where $C=\lfloor \Theta/2\rfloor$, $\Theta$ is the width and height of the filter, $F$ is the number of input channels, $g$ is the number of output channels, $b$ is the bias and $\varphi$ is an activation function.
As activation function, we choose ReLU~\cite{NH2010} which bypasses the problem of vanishing gradients.
For training this architecture, weights ${\bm w}$ are optimized to minimize the loss function between estimated and reference data ${\bm q}_{\rm Ref}$:
\begin{equation}
    {\bm w}={\rm argmin}_{\bm w}||{\bm q}^{(l_{\rm max})}-{\bm q}_{\rm Ref}||_2,
    \label{eq2}
\end{equation}
where ${\bm q}^{(l_{\rm max})}$ is an output of CNN at the last layer $l_{\rm max}$.
We use the Adam optimizer~\cite{kingma2014} for initial training before deploying SWAG.

\subsubsection{Probabilistic neural network}
\label{sec:PNN}

For comparison to the present SWAG-based UQ which assesses epistemic uncertainty, we consider the probabilistic neural network (PNN)~\cite{MDN1994}, which estimates data uncertainty.
As mentioned in introduction, PNNs have been applied to reduced-order modeling and global-field reconstruction of fluid flows~\cite{MFRFT2020}.
The aforementioned neural networks such as MLP and CNN are generally trained to obtain optimized weights ${\bm w}$ by minimizing a loss function.
However, the output obtained through this deterministic fashion cannot provide information about the epistemic uncertainty.
In the PNN formulation, we are able to obtain the probability distribution of the estimations $p({\bm q}_{\rm out}|{\bm q}_{\rm in})$.
The PNN-based mapping can be expressed as $\mathcal{F}: {\bm q}_{\rm in} \rightarrow (\pi_1, \mu_1, \sigma_1, \pi_2, \mu_2, \sigma_2, ...,  \pi_S, \mu_S, \sigma_S)$, where $\pi_i$ is the mixing probability for each Gaussian component satisfying the condition $\sum_{i=1}^{m}\pi_i = 1$, whereas the mean $\mu$ and the standard deviation $\sigma$ determine a Gaussian probability distribution function $\mathcal{N}(\mu,\sigma)$. 
The distribution function of PNN can be assumed as a linear combination of these Gaussian components such that:
\begin{equation}\label{eq:GMM}
    p({\bm q}_{\rm out} | {\bm q}_{\rm in}) = \sum_{i=1}^{m}\pi_i ({\bm x}) \mathcal{N}(\mu_i ({\bm q}_{\rm in}),\sigma_i ({\bm q}_{\rm in})).
\end{equation}
In the present study, the value of $m$ is set to 1 following our previous work~\cite{MFRFT2020}.
Note that care should be taken in choosing the loss function to utilize the full distribution of the estimation because the PNN outputs a distribution of the estimation $p({\bm q}_{\rm out}|{\bm q}_{\rm in})$ instead of a target variable of ${\bm q}_{\rm out}$.
Hence, the cost function $E$ for the PNN ${\mathcal F}({\bm q}_{\rm in};{\bm w})$ is given with regard to the average log-likelihood $\mathcal{L}$ such that:
\begin{equation}\label{eq:negloglike}
\begin{gathered}
    {\bm w}={\rm argmin}_{\bm w}[\mathcal{E}],\quad \text{where} \quad \mathcal{E} \equiv -\log \mathcal{L} = -\sum_{k=1}^{K} p({q}_{{\rm out},k,p}|{q_{{\rm in},k}}) \log p({q}_{{\rm out},k,t}),
\end{gathered}
\end{equation}
where $k$ indicates each data point in training data and $K$ represents the number of training samples.
Note that maximizing the likelihood can be regarded as the equivalent operation to minimizing the cross entropy $H(p({\bm q}_{\rm out}|{\bm q}_{\rm in}), p({\bm q}_{\rm out}))$. Our comparison of SWAG and PNN is performed for the cylinder-wake example.


\section{Results and discussion}
\label{sec:results}
\subsection{Global field reconstruction from sparse sensor measurements}
\label{sec:sens2field}

\subsubsection{Example 1-1: two-dimensional cylinder wake at various Reynolds numbers}
\label{sec:cylinder}

\begin{figure}
  \centering
    \includegraphics[width=\textwidth]{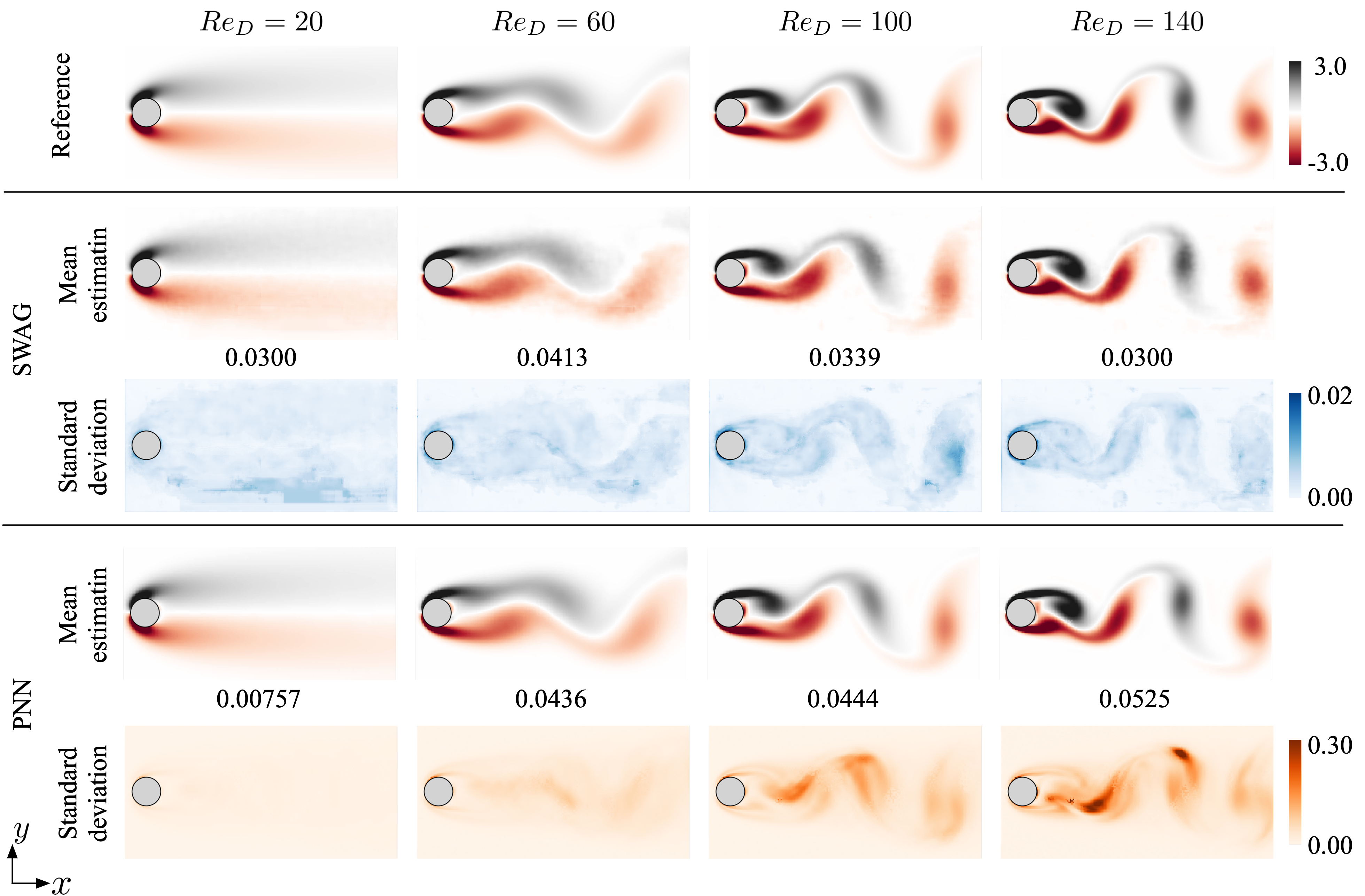}
    \caption{Mean and standard deviation of the estimated cylinder wake with SWAG and PNN.
    Sixteen sets of the weights are sampled from the approximated Gaussian distribution of the weights.
    The distribution is approximated with the weights obtained through SGD with its constant learning rate of $1.0\times10^{-3}$.
    The values underneath the contours represent the $L_2$ error norm, $||{\bm q}_{\rm Ref}-{\bm y}_{\rm SWAG}||_2/||{\bm q}_{\rm Ref}||_2$.}
    \label{fig:cylinder}
\end{figure}

\begin{figure}[t]
  \centering
    \includegraphics[width=\textwidth]{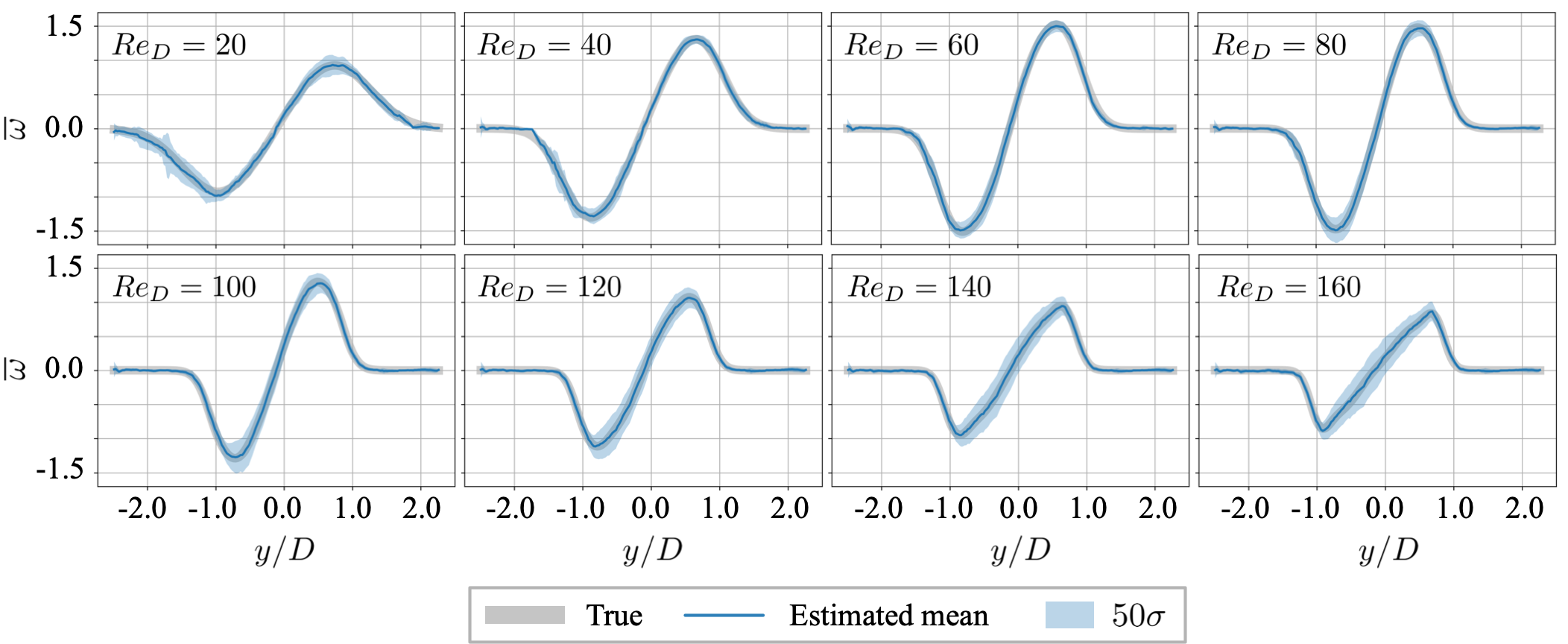}
    \caption{Time-averaged vorticity profile estimated via SWAG-based approximation at $x/D=11$ for each Reynolds number. The shaded region corresponds to the standard deviation over the epochs.}
    \label{fig:cy_avgvort}
\end{figure}

\begin{figure}
  \centering
    \includegraphics[width=\textwidth]{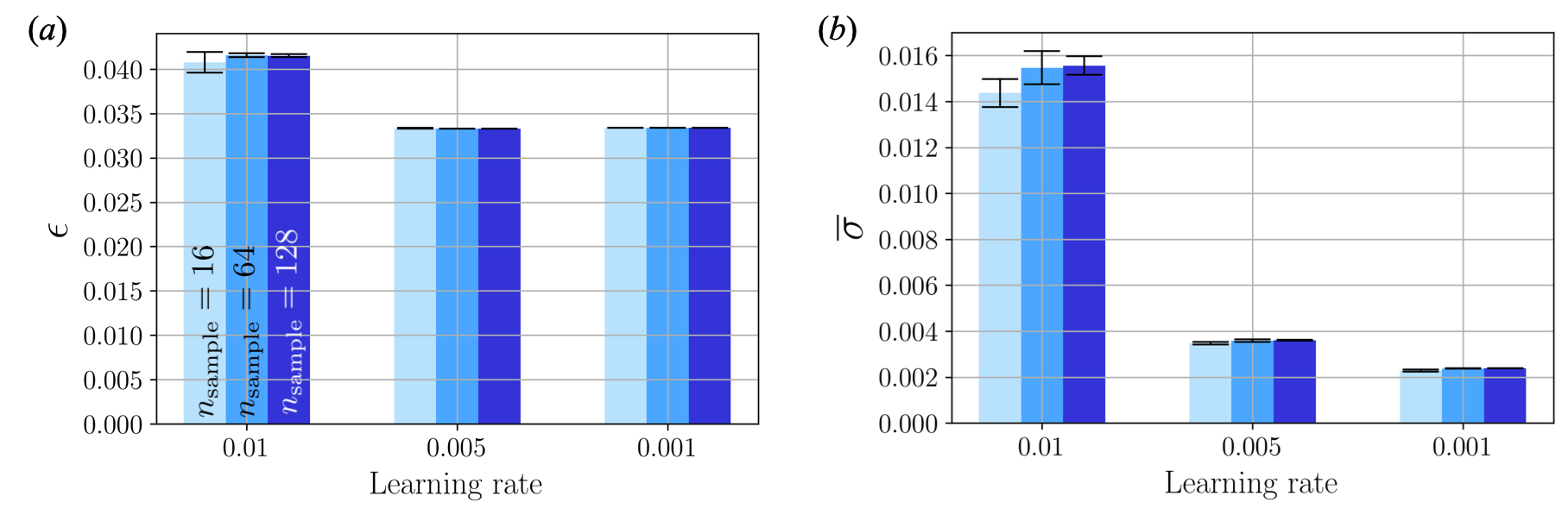}
    \caption{Dependence of $(a)$ the estimation error and $(b)$ the average standard deviation on the choice of the learning rate and the number of the weight samplings of SWAG. The errorbar indicates the standard deviation among three-fold weight samplings from the distribution.}
    \label{fig:cy_nsample}
\end{figure}

We first consider the flow around a cylinder at $Re_D=\{20,40,60,80,100,120,140,160\}$ \cite{HFMF2020b} to investigate the capability of SWAG-based UQ for flows at various low Reynolds numbers.
The datasets are generated with a two-dimensional direct numerical simulation (DNS) by numerically solving the incompressible Navier--Stokes equations~\cite{kor2017}.
We use 3000 snapshots of the vorticity field $\omega$ at each Reynolds number for training.
The aim of the CNN-MLP model is to estimate the wake-vorticity field ${\omega}$ from five vorticity sensors ${\bm s}$ located on the cylinder surface.
Since a single model is trained for flows at eight different Reynolds numbers, the Reynolds number itself is also utilized as the supplemental input information for the model~\cite{MFZNF2021}, such that the input of the present model is ${\bm q}_{\rm in}=\{{\bm s},Re_D\}\in {\mathbb R}^6$.

The estimated results with SWAG at various Reynolds numbers are shown in figure~\ref{fig:cylinder}.
The mean estimation of SWAG shows a reasonable agreement with the reference data.
Notably, regions of large standard deviation show similar structures when compared to the wakes for each Reynolds number, which suggests that the epistemic uncertainty correlates with the unsteadiness of the flow field.
Similar observations can be made from the averaged vorticity distribution in figure~\ref{fig:cy_avgvort}.
For comparison, we also display the result from probabilistic neural network (PNN)~\cite{MFRFT2020} in figure~\ref{fig:cylinder}.
Note that we subsample the flow field to reduce the size of the data to a quarter of the original DNS (i.e., coarsened in each direction) because PNN is configured in a fully connected manner, which is unsuitable for handling high-dimensional data.
Similar to the reconstruction with the SWAG, the mean estimations are in agreement with the reference data for the all considered Reynolds numbers.
The distribution of standard deviation obtained through the PNN is also similar to that of the SWAG.
Although both UQ methods capture the increase of uncertainty with $Re_D$, which associated with unsteadiness of wakes, implications of them are different with each other.
While PNN concentrates on a probability distribution of the given data, a standard deviation of SWAG reflects the result of weight fluctuations around the end of the training.
The standard-deviation estimates for both these techniques are complementary, with SWAG capturing the effect of (potentially) insufficient data and PNNs being useful for characterizing the inherent variation in the training data itself.

A comparison of computational costs between the PNN and the SWAG-based method is summarized in table~\ref{tab:comp_cost}.
For this comparison, the training for each case is performed with 100 snapshots of cylinder wake at $Re_D=100$ using a graphical processing unit (NVIDIA GeForce RTX 2080 Ti).
Regarding the computational cost for training, PNN requires approximately seven folds longer computational time than the pre-training of MLP-CNN model.
This is because a PNN is built in a fully-connected manner, which generally requires significantly larger amounts of the weights than a CNN model.
Although the post-training of SWAG with SGD requires a longer computational time than the pre-training, it is still shorter than that of the PNN.
Note that the computational cost for the evaluation of the SWAG-based method is longer than that for PNN since multiple estimations, i.e., $n=16$ in this case, are required in obtaining the ensemble result.
Therefore, the MLP-CNN model with SWAG-based learning requires approximately $75\%$ of the computational cost from PNN in an overall sense, which indicates that the SWAG-based uncertainty quantification can be performed at a reasonable computational cost.
However, we note that this figure may be affected by the choice of the model hyperparameters as well as the architecture.

\begin{table}
\centering
\caption{Comparison of computational costs between PNN and SWAG-based learning with a graphical processing unit (NVIDIA GeForce RTX 2080 Ti).
As for the evaluation time for SWAG, the estimation is performed sixteen folds to obtain a multiple estimation where a single evaluation requires $0.849$ seconds.}
\label{tab:comp_cost}
\vspace{3mm}
\begin{tabular}{ccccc}
    & & Training time (s) & Evaluation time (s)  & Total (s)\\\hline
PNN & & 1245              & 1.03                 & 1246\\\hline
\multirow{2}{*}{SWAG} & pre-training & 184 & -- & \multirow{2}{*}{933}\\
    & post-training & 735 & 0.849~s/fold~$\times$~16~folds~=~13.6 & \\\hline
\end{tabular}
\end{table}

Let us then investigate the sensitivity regarding the hyperparameters of SWAG in figure~\ref{fig:cy_nsample}.
We focus on two crucial parameters which may affect on the results of SWAG: 
\begin{enumerate}
    \item \underline{Learning rate $\eta$ of SGD during SWAG sampling:}\\
    We can expect that SGD iterations with larger learning rates would explore a larger region of the loss surface, which may lead to an inaccurate approximation of the weight distributions. 
    This would be due to the sampling of low-performing weights far away from the `good-regions' of the loss surface.
    On the other hand, with a very small learning rate, it is difficult to sample weights that are sufficiently diverse while exhibiting good performance in the function approximation.
    Due to this trade-off at stake here, a proper selection of the learning rate $\eta$ is crucial for successfully performing SWAG-based UQ.
    \item \underline{Frequency of weight sampling from the Gaussian distribution --$n$ in equation~(\ref{eq:5})--:}\\
    After approximating a distribution for the weights, we can sample multiple sets of the weights and create multiple models, as expressed in equation~(\ref{eq:5}). Since a mean and a standard deviation are obtained through those multiple estimations, we examine the convergence of the results with respect to the number of samples.
\end{enumerate}

We consider the combinations of three different learning rates $\eta=\{0.001, 0.005, 0.01\}$ and three different numbers of samples $n=\{16, 64, 128\}$ for the assessment in figure~\ref{fig:cy_nsample}. The error bars in this figure indicate three repetitions of the SWAG experiment with different random seeds.
Focusing on the influence of the learning rate in figure~\ref{fig:cy_nsample}$(a)$, the variation of the $L_2$ error is relatively low, i.e., $\epsilon \approx 0.03-0.04$, among the considered learning rates.
On the other hand, a significant difference can be observed in a comparison of the standard deviation averaged in space and time $\overline{\sigma}$, as presented in figure~\ref{fig:cy_nsample}$(b)$.
With the large learning rate, i.e., $\eta=0.01$, the standard deviation is larger than that with the smaller rates, i.e., $\eta=\{0.005, 0.001\}$.
This is likely because the SGD with the large learning rate explores a wider area of the loss surface as discussed above, which leads to the larger covariances for the fit distribution.
In contrast, there is no significant difference for the number of samplings in terms of both the $L_2$ error and the standard deviation, as shown in figure~\ref{fig:cy_nsample}.
This indicates that some form of convergence has been reached and we may use 16 samples for our UQ estimation.

\subsubsection{Example 1-2: DayMET data set}
\label{sec:daymet}

\begin{figure}[t]
  \centering
    \includegraphics[width=\textwidth]{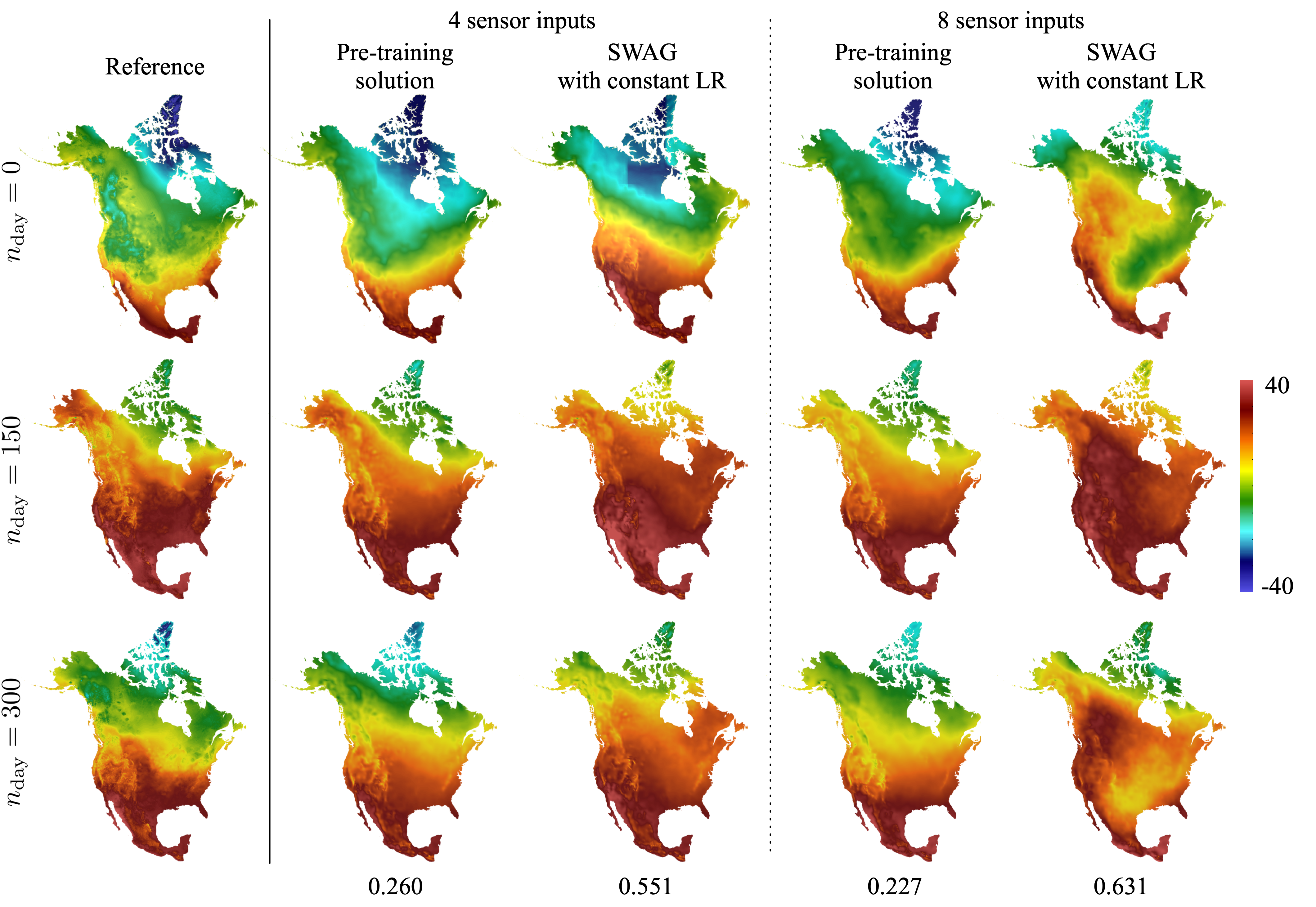}
    \caption{Applications of SWAG to the DayMET dataset.
    Days are represented with $n_{\rm day}$, where $n_{\rm day}=0$ is January 1st and $n_{\rm day}=364$ is December 31st.
    Results of the pre-training solution, SWAG with constant learning rate and SWAG with cyclical learning rate are presented, for both 4 and 8 sensors as input.
    The values underneath each figure represents the instantaneous $L_2$ error norm for each day.}
    \label{fig:daymet_result}
\end{figure}

\begin{figure}[t]
  \centering
    \includegraphics[width=\textwidth]{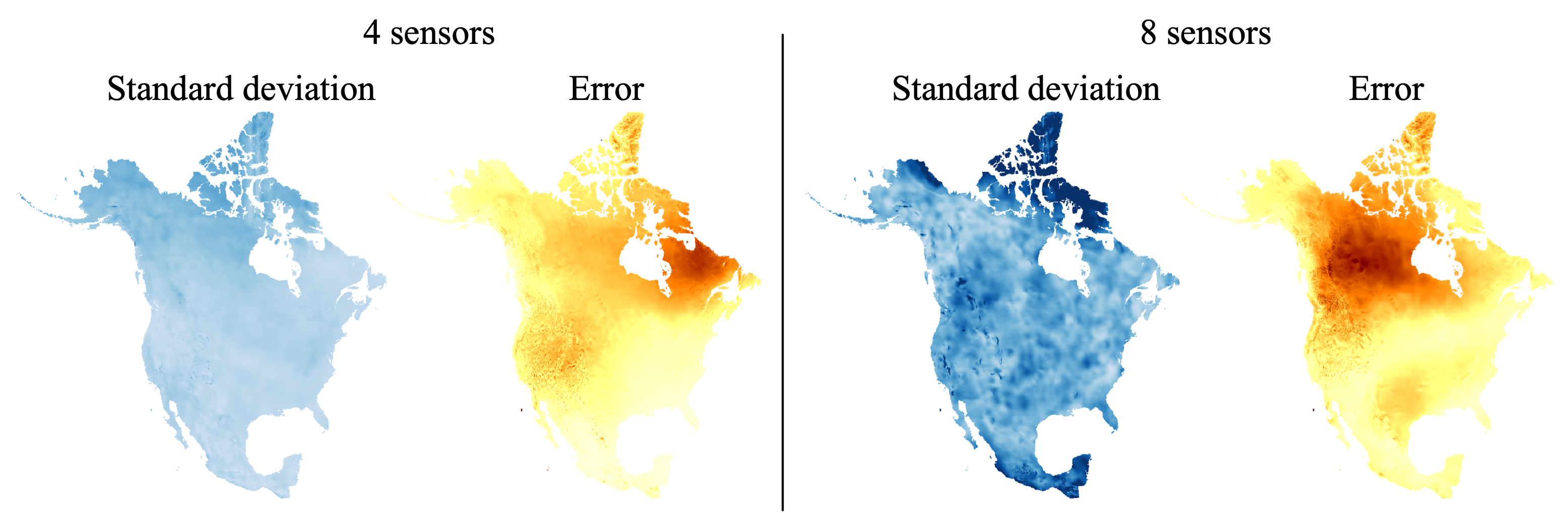}
    \caption{{Time-averaged fields of estimated standard deviation and absolute error of mean estimation obtained through the present SWAG with the constant learning rate.
    The standard deviation is colored in the range of 0.0 to 1.5 while the absolute error is in the range of 0 to 30 with darker color representing larger value.}
    }
    \label{fig:daymet_std-err}
\end{figure}

\begin{figure}
  \centering
    \includegraphics[width=\textwidth]{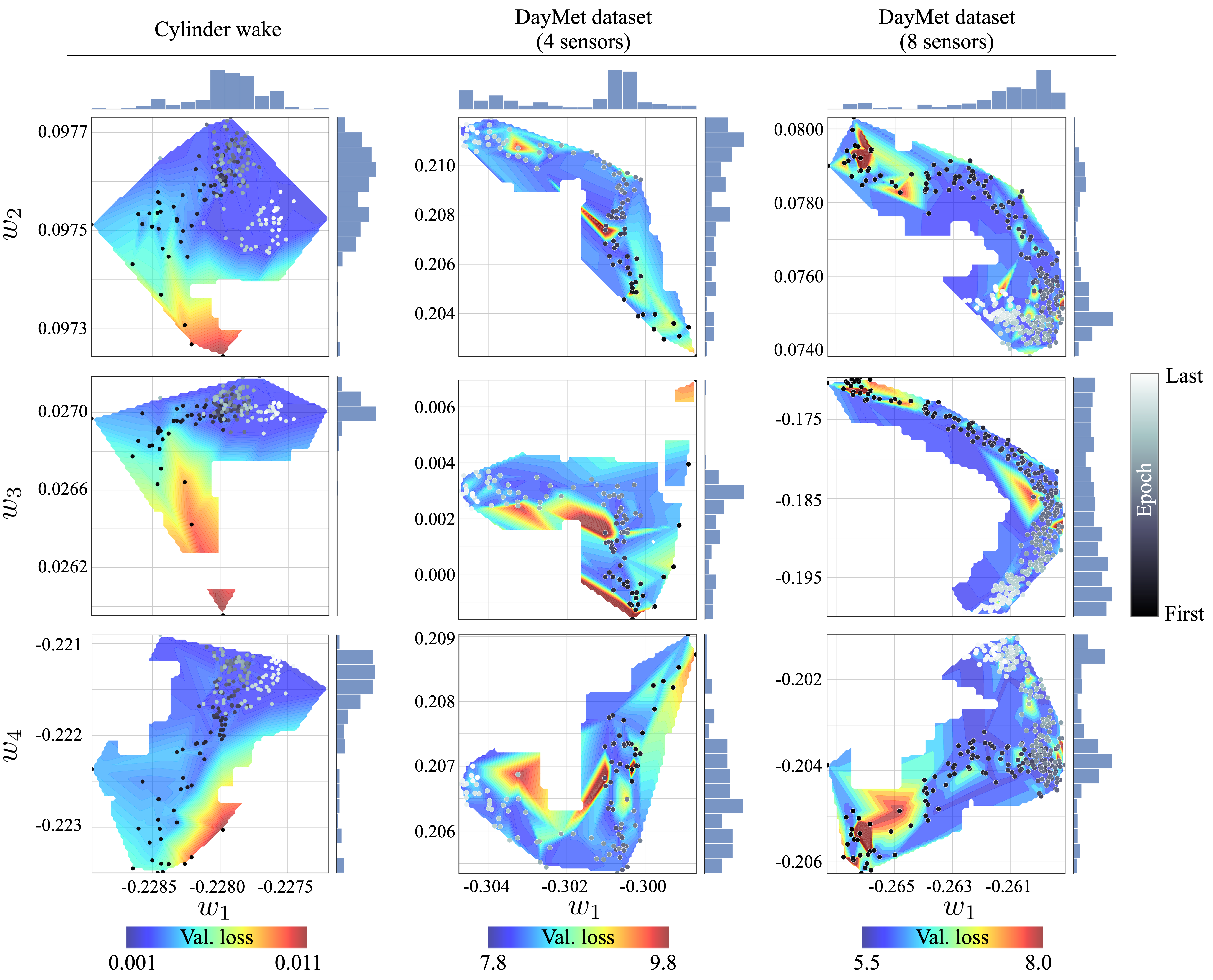}
    \caption{Loss surface of weights inside the present models for the cylinder-wake and the DayMET datasets.
    {Training history of the weights is expressed with the circle markers.
    The darkest-colored markers represent the first epoch and the lightest-colored ones represent the last epoch.}
    The histogram given for each axis on each contour represents the distribution of the weights.}
    \label{fig:loss_surface}
\end{figure}

\begin{figure}
  \centering
    \includegraphics[width=0.9\textwidth]{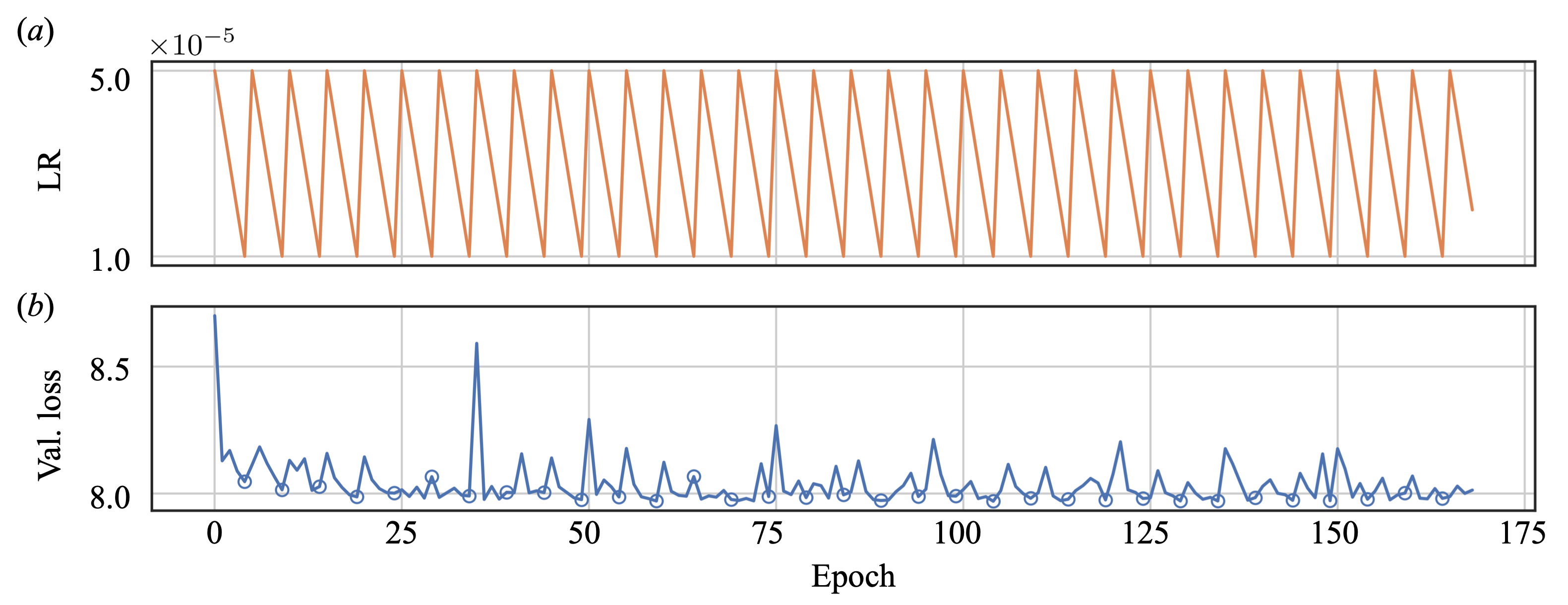}
    \caption{An example of the cyclic-learning-rate schedule. $(a)$ History of the leaning rate and $(b)$ validation loss for the DayMET data reconstruction with four sensors. Blue circles indicate the epochs with a minimum learning rate.}
    \label{fig:cyclic_concpet}
\end{figure}

\begin{figure}
  \centering
    \includegraphics[width=\textwidth]{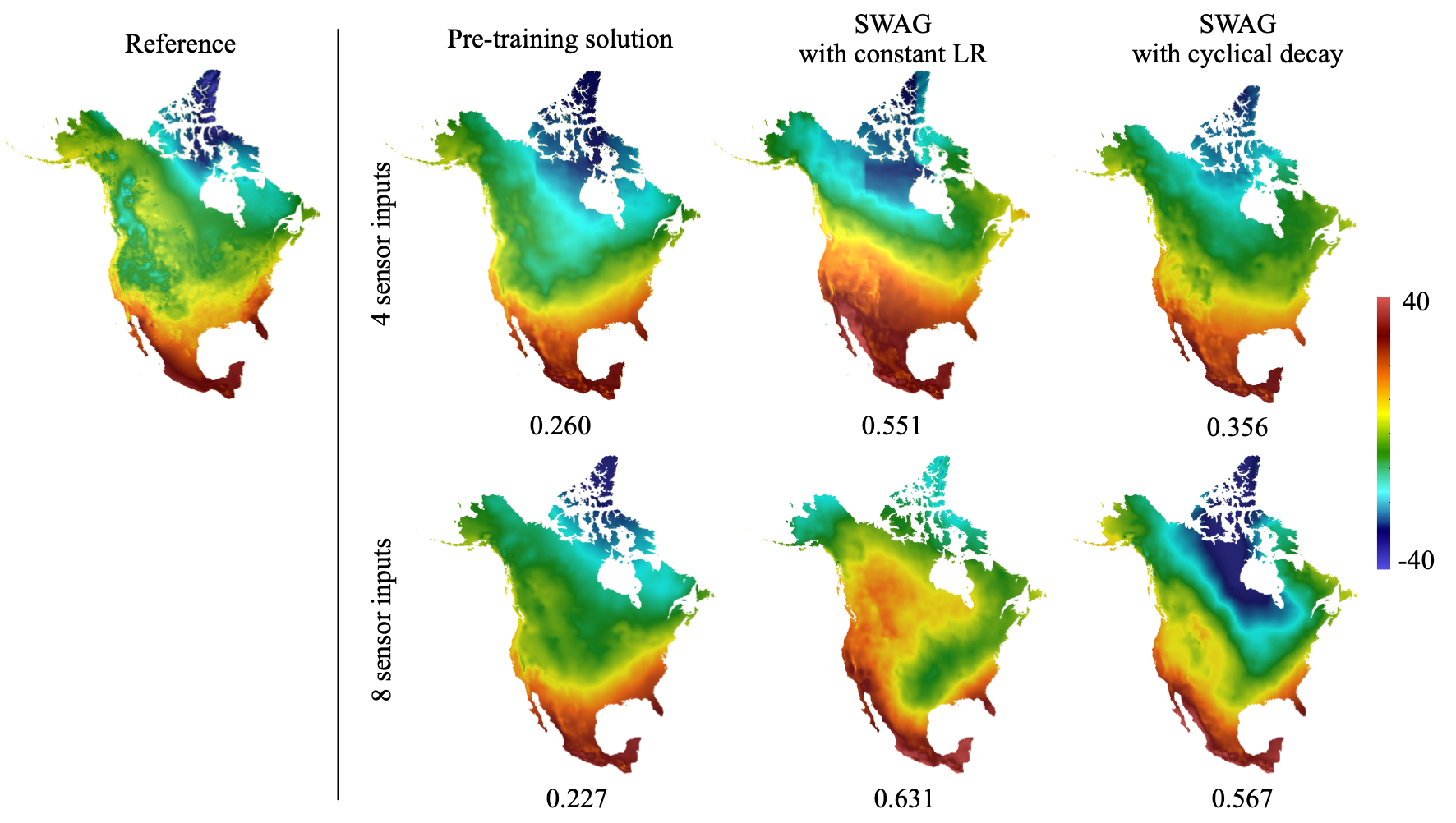}
    \caption{Dependence of the SWAG performance on the use of constant learning rate and cyclic-learning-rate schedule for the DayMET dataset.
    The estimated fields at $n_{\rm day}=0$ are shown. The values underneath each fields represent an $L_2$ error norms calculated over whole dataset.}
    \label{fig:daymet_cyclic}
\end{figure}

\begin{figure}[t]
  \centering
    \includegraphics[width=0.85\textwidth]{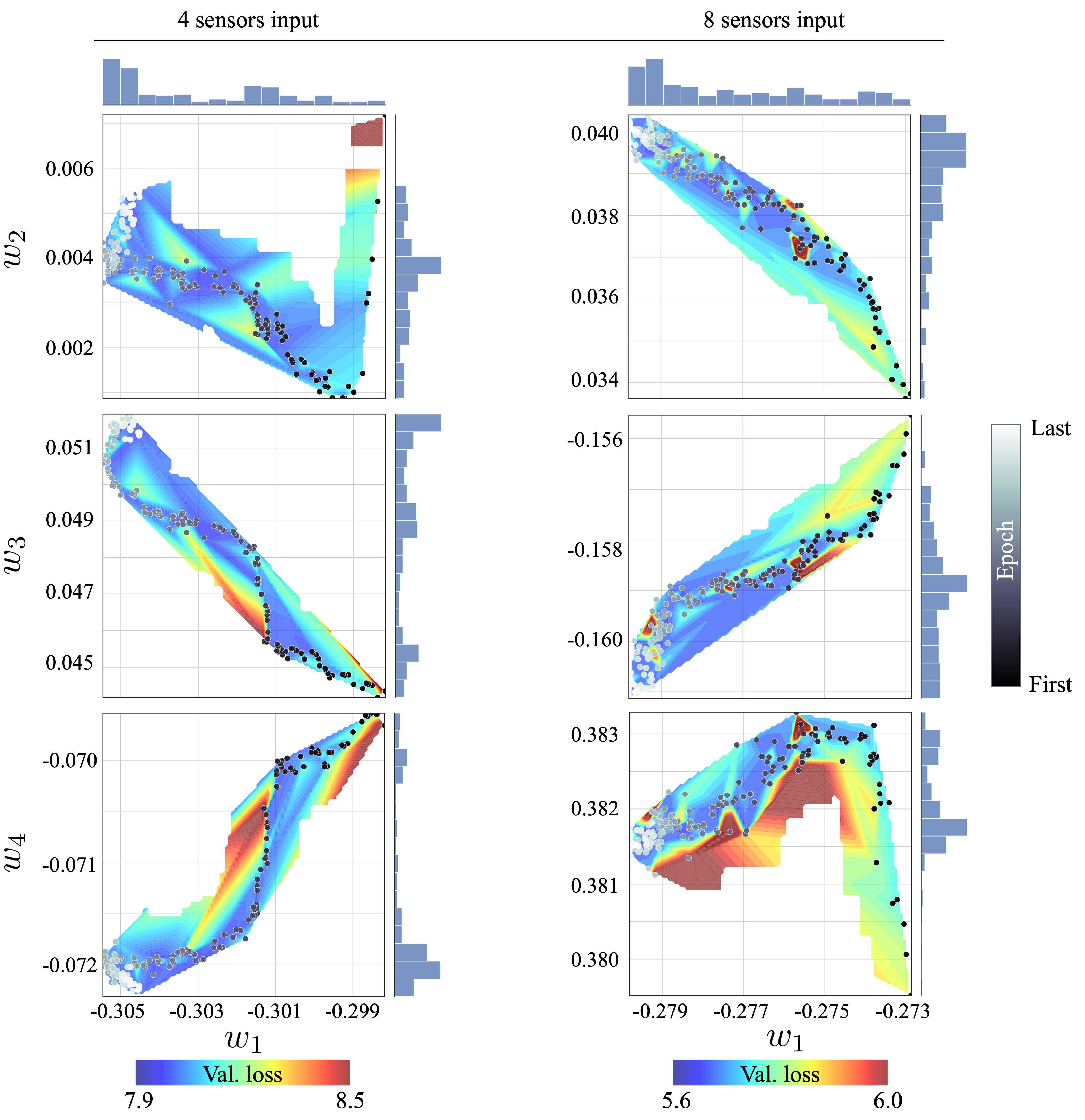}
    \caption{Loss surface of weights inside the model trained with cyclical learning rate schedule for the DayMET dataset.
    {Training history of the weights expressed with the circle markers.
    The darkest-colored markers represent the first epoch and the lightest-colored markers represent the last epoch.}
    The histogram given for each axis on each contour represents the distribution of the weights.}
    \label{fig:loss_surface_cyclical}
\end{figure}

\begin{figure}
  \centering
    \includegraphics[width=\textwidth]{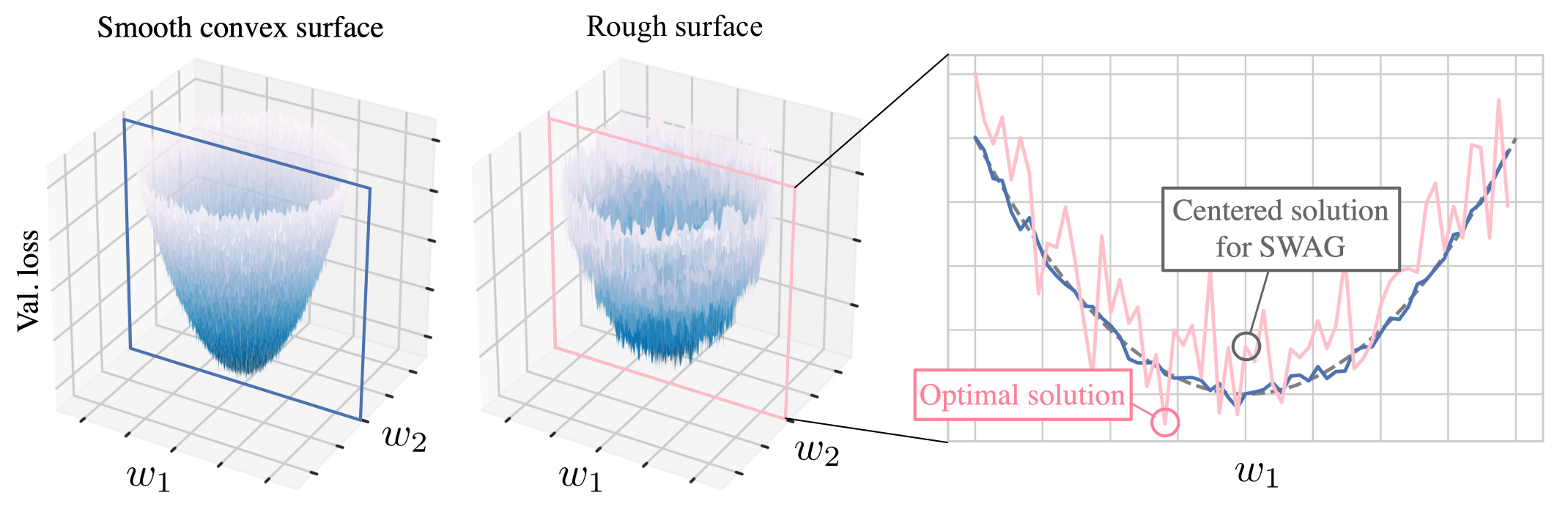}
    \caption{{Schematic of smooth and rough error surfaces over two arbitrary weights.
    The transition of the error near the optimal point is shown on the right side.
    Over the rough surface (pink line), the centered solution may not be located at an accurate position since multiple possible solutions may be observed due to the SGD iterations with the cyclical learning rate schedule.}}
    \label{fig:loss_surface_schem}
\end{figure}

Although we demonstrated the applicability of SWAG for unsteady laminar wakes above, we should note that SWAG has a limitation associated with a multimodality of loss surface caused by a problem setting and data sets that users handle.
To examine the limitation of the present scheme, we consider the DayMET data set (\url{https://daymet.ornl.gov/}) which provides the maximum daily temperature field across North America in the period 2000--2016.
The CNN-MLP model aims to reconstruct a whole temperature field from four or eight randomly placed temperature-sensor measurements, which can be regarded as a difficult problem due to the high dimensionality of the system.
We use the first 15 years for training, while the remaining is used for testing purposes.
Hereafter, we do not consider the use of a standard PNN due to the computational limitation associated with its fully-connected structure, as mentioned earlier.
Note that convolutional PNNs may be devised, with output channels devoted to predicted mean and variance estimates at each point of a structured grid; however, we leave this to a future study.

The estimated temperature fields for representative snapshots are shown in figure~\ref{fig:daymet_result}.
The best set of weights obtained through pre-training with the Adam optimizer can capture a seasonal temperature variation from four or eight sensors with relative $L_2$ error norms of approximately 20--25\%.
However, by applying the SWAG-based approximation, the estimation is worse than that with the original model with the $L_2$ error norms over 50\% for both numbers of sensors.
Moreover, despite the fact that the pre-training solution with eight sensors outperforms that with four sensors, the SWAG-based approximation exhibits the opposite trend, with higher error in the case with larger number of sensors.

{The limitation in applying the present method to this particular high-dimensional system can also be observed through the comparison of time-averaged fields of estimated standard deviation and absolute error of mean estimation shown in figure~\ref{fig:daymet_std-err}.
For both input attributes, the high concentration regions of standard deviation and error are different from each other.
This trend indicates that the present method cannot observe `well-calibrated' epistemic uncertainty, as also discussed in section~\ref{sec:sqcy}.}
These behaviors imply that the loss surface of this problem setting could not be properly approximated with a Gaussian distribution.

To clarify an influence of the explored loss surface on the approximation of the weight distribution, we visualize the loss surface of each machine-learning model in figure~\ref{fig:loss_surface}.
We randomly choose a weight $w_1$ and plot its history over training epochs against three other randomly-chosen weights $w_2$, $w_3$, and $w_4$.
For comparison, the same analysis is also applied to the case with the cylinder wakes as an example of the successful Gaussian approximation.
{The training history of the weights is expressed with the circle markers.
The darkest-colored markers represent the first epoch and the lightest-colored ones represent the last epoch.}
The weights basically distribute around the region with a relatively small validation loss, i.e., blue region in figure~\ref{fig:loss_surface}.
A histogram of each weight shown aside the loss surface also indicates that the distribution of the weights concentrates on these regions.
Moreover, we also find that a stability of the training procedure can be considered as an important factor for successfully performing SWAG.
Since the validation loss gradually decreases and converges, a loss surface results in a `smooth' surface, as shown in figure~\ref{fig:loss_surface}.
{In other words, a smooth transition of validation errors over the weight surface can be observed for this particular example.}
This behavior helps the accurate approximation of a Gaussian distribution of the weights and leads to a scalable epistemic uncertainty quantification.

In contrast, the training procedure becomes significantly unstable with the DayMET dataset and the loss surface exhibits multimodality, which was not the case in the cylinder wake.
Especially with eight sensors, the validation loss fluctuates even at the end of the training procedure and the weights does not converge with a small validation loss.
{Compared to the error surface of the cylinder case, the surfaces of the DayMET data clearly exhibit a high gradient of error over the weights ${\rm d}\epsilon/{\rm d}{\bm w}$.}
Furthermore, as observed from the loss surface among $w_1$ versus $w_3$ of the case with eight sensors, the histogram of the weights tends to be flatter than that of the cylinder case, which may lead to an inappropriate approximation of the central solution.
These behaviors of the weight distribution hinder the accurate approximation of posterior distribution and generate a set of weights which cannot even perform the mean estimation properly.

{
To quantitatively account for the entire weights in the model, we also evaluate a gradient of the error over weight surfaces $\partial \epsilon/\partial{\bm w}$ for every weight combination.
A maximum gradient of the error is calculated with every weight combination, and we then take the average over the entire set of the weights.
For the model trained with the cylinder wake example, the averaged maximum loss gradient is $2.93\times10^2$ whereas that of the model trained with eight sensor measurements of the DayMET dataset is $1.36\times10^7$.
This indicates that the model trained with the DayMET dataset is more likely to be associated with the multimodal error surfaces with sharp gradients, and is therefore not amenable to a convex approximation near the optima.
Thus, it is quantitatively clarified that the Gaussian approximation of the weights was unsuccessful for the case with the DayMET dataset compared to the cylinder wake case.}
This is a problem with SWAG when handling a high-dimensional nonlinear problem.

To overcome this issue, we consider applying a cyclical-learning-rate schedule~\cite{SWAarxiv2019}, which was designed for the purpose of improving SWAG, instead of SGD having a constant learning rate.
The learning rate linearly decreases from $a_1$ to $a_2$, where $a_1>a_2$, with its cycle step length with $c$.
We set the maximum learning rate to $5.0\times10^{-5}$ and its minimum to $1.0\times10^{-5}$, as shown in figure~\ref{fig:cyclic_concpet}.
As the training process proceeds, the learning rate decreases from $a_1$ to $a_2$ within five epochs and jumps to $a_1$ at the next epoch.
A history of validation loss for the case with four sensors is also shown under the history of learning rate in figure~\ref{fig:cyclic_concpet}$(b)$.
The weights at the blue circled epochs in figure~\ref{fig:cyclic_concpet}$(b)$, where the iteration is performed with a minimum learning rate, are only adopted to approximate a posterior distribution of the weights because we can expect that it results in a small validation loss.

The estimated results with cyclical-learning-rate schedule are summarized in figure~\ref{fig:daymet_cyclic}.
For the case with four sensors, the result with cyclical learning rate outperforms that with constant one, with an $L_2$ norm approximately 20\% lower. Although the estimation error is still larger than the pre-training solution, the use of cyclical learning rate can be one of the candidates to mitigate the issue of unstable training procedure and improve the applicability of SWAG-based UQ on highly-sensitive problem settings. 

However, it should be noted that the use of cyclical-learning-rate decaying does not show such a significant improvement for the case with eight sensors.
This is likely because of the distribution of weights, as presented in figure~\ref{fig:loss_surface_cyclical}.
The distribution of weights for the case with four sensors concentrates around the region with smaller validation loss at the end of the training, which leads to a better determination of a centered solution.
On the other hand, a rough surface with sharp gradients that does not suit a convex approximation can be observed for the case with eight sensors.

{We present a simplified schematic about the differences between the `smooth' and `rough' surfaces in figure~\ref{fig:loss_surface_schem}.
Over the smooth convex-shaped error surface (e.g., cylinder wake case and cyclical-learning-rate decaying for the DayMET dataset with four sensor measurements), the Gaussian approximation can be accurately performed from its error transition, as represented in a blue line.
However, with the rough surface with sharp gradients that does not suit a convex approximation, represented as a pink line, the centered solution of the weight may not fit around the optimal solution but rather results in an inappropriate high-error position.
Since the SGD with cyclical learning rate explores various optimal valleys of the multimodal loss surface, the centered solution of collected weights cannot find a better position and causes the inaccurate approximation of the weight distribution.}

Furthermore, the choice of parameters for cyclical learning rate significantly influences the result of SWAG.
In addition to the baseline setup (SWAG with cyclical decay in figure~\ref{fig:daymet_cyclic}), three cases are used to examine the robustness of the algorithm, as summarized in table~\ref{tab:cyclic_param}.
For these cases, three parameters of cyclical learning rate are considered: (i) a maximum learning rate $\eta_{\rm max}$, (ii) a minimum learning rate $\eta_{\rm min}$, and (iii) a cycle length $c$.
Although an improvement in the estimation error can be observed for all cases compared to the result with constant learning rate, the results are significantly influenced by the parameter settings with their largest difference in the $L_2$ error norms of approximately 15\%.
The best case for each input configuration is also different to each other.
Thus, although a customization of an optimization method, e.g., cyclical-learning-rate scheduling, can be one of the candidates to mitigate the issue, care should be taken depending on problem settings and parameter choices for optimization methods.

\begin{table}
\centering
\caption{Parametric influence on cyclical-learning-rate schedule for the DayMET data estimation. The $L_2$ error norm of the estimated fields $\epsilon=||{\bm q}_{\rm Ref}-{\bm y}_{\rm SWAG}||_2/||{\bm q}_{\rm Ref}||_2$ are summarized.}
\label{tab:cyclic_param}
\begin{tabular}{cccccc}
          & $\eta_{\rm max}$   & $\eta_{\rm min}$   & $c$ & $\epsilon$ (4 sensors) & $\epsilon$ (8 sensors)\\\hline
Constant learning rate & \multicolumn{2}{c}{$\eta=1.0\times10^{-5}$} & -- & 0.551 & 0.631\\
Baseline & $5.0\times10^{-5}$ & $1.0\times10^{-5}$ & 5   & 0.356                & 0.567\\
Case 1   & $5.0\times10^{-5}$ & $1.0\times10^{-6}$ & 5   & 0.505                & 0.414\\
Case 2   & $9.0\times10^{-5}$ & $1.0\times10^{-5}$ & 5   & 0.463                & 0.542\\
Case 3   & $5.0\times10^{-5}$ & $1.0\times10^{-5}$ & 9   & 0.382                & 0.546\\\hline
\end{tabular}
\end{table}

\subsection{Estimation of far-field state from two-dimensional fields}
\label{sec:sec2sec}

\subsubsection{Example 2-1: Three-dimensional square-cylinder wake at $Re_D=300$}
\label{sec:sqcy}

\begin{figure}[t]
  \centering
  \includegraphics[width=\textwidth]{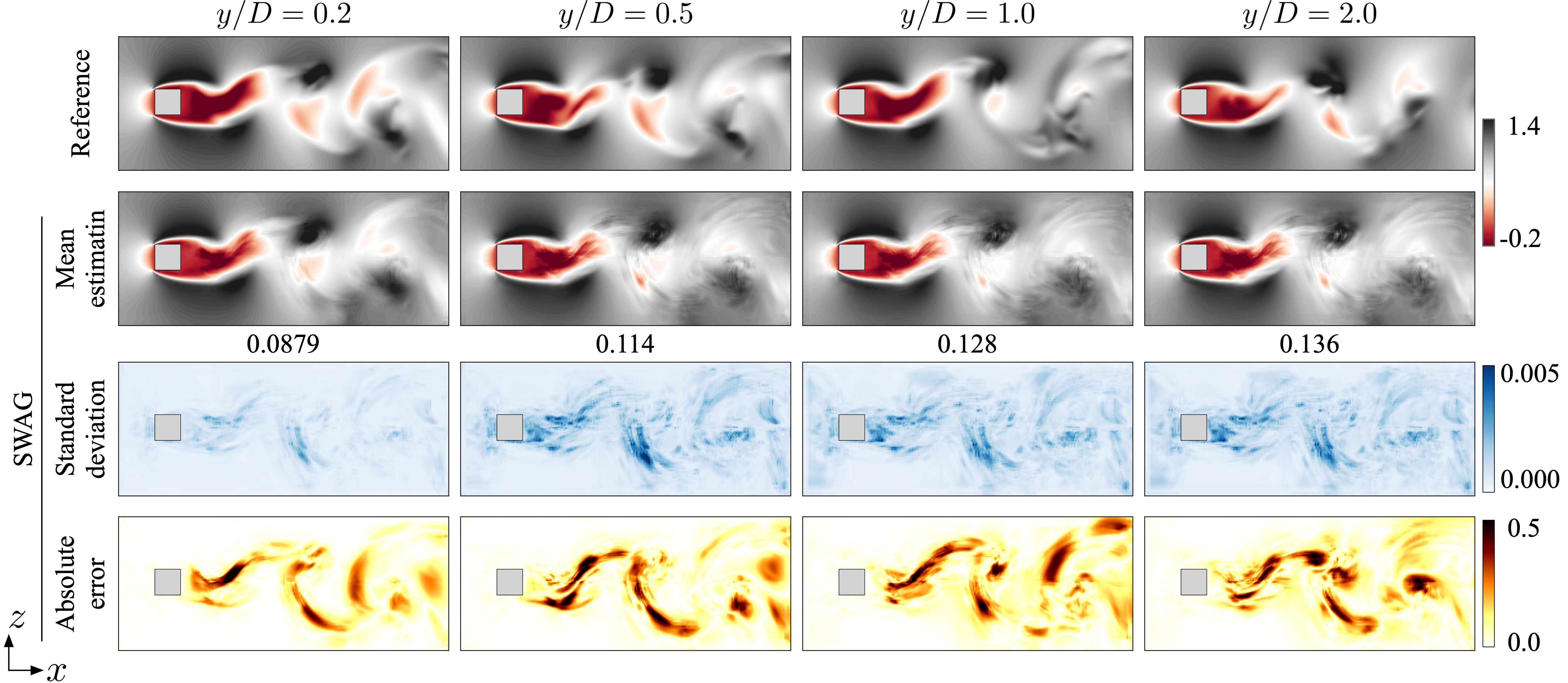}
  \caption{Two-dimensional $x-z$ sections of streamwise velocity in a square-cylinder wake at $y/D=0.2-2.0$ estimated from that at $y/D=0.0$.
  {Fields of mean estimations, standard deviations, and absolute errors obtained through the present SWAG are depicted.}
  The values underneath the estimated field represent the $L_2$ error norm.}
  \label{fig:sqcy}
\end{figure}

We now apply the SWAG method to far-field state estimation from sectional data using convolutional neural networks (CNNs).
As the first example in this section, let us consider a three-dimensional square cylinder wake at $Re_D=300$~\cite{MFF2021}.
A direct numerical simulation (DNS) is utilized to prepare the dataset.
The immersed square cylinder is expressed with a penalization term~\cite{volumePenal1994} in the incompressible Navier--Stokes equations, 
\begin{flalign}
    &{\bm{\nabla}} \cdot {\bm u}=0, \\
    &\frac{\partial {\bm u}}{\partial t} + {\bm{\nabla}} \cdot \left({\bm{u}}{\bm{u}}\right)=-{\bm{\nabla}} p + {{ Re}^{-1}_D}{\bm{\nabla}}^2\bm{u}+\lambda \chi\left({\bm u}_b-{\bm u}\right),
\end{flalign}
where $\lambda$, $\chi$, and ${\bm u}_b$ are the penalty parameter, the mask value ($0$ outside and $1$ outside and inside a body) and the velocity inside the object (which is zero), respectively.
The velocity vector ${\bm u}=\{u,v,w\}$ and pressure $p$ are non-dimensionalized by the fluid density $\rho$, the length of one side of the square cylinder $D$, and the freestream velocity $U_\infty$.
The computational domain has dimensions $\left(L_x, L_y, L_z\right)=\left(20D, 4D, 20D\right)$ and the time step is set to $\Delta t=2.5\times10^{-3}$.
We focus on the region around the body with size: $\left(12.8D, 4D, 4D\right)$, with the following number of grid points: $(\hat{n}_x, \hat{n}_y, \hat{n}_z)=(256, 160, 128)$.
For more detailed setup of the data, we refer readers to Morimoto et al.~\cite{MFF2021}.
We use 2,000 snapshots for both pre-training and the second training with SGD.
The aim of the present CNN model is to estimate four $x-z$ cross sectional velocity fields at $y/D=\{0.2,0.5,1.0,2.0\}$ from a velocity field at $y/D=0.0$.
A streamwise velocity field $u$ is utilized for both the input and output attributes.

The application of SWAG-based estimation and UQ for the square-cylinder wake is summarized in figure~\ref{fig:sqcy}.
As can be expected, the estimation error of the section at $y/D=0.2$ (the one closest to the input section) reports the lowest error among the covered sections.
Similarly, the standard deviation is also smaller at $y/D=0.2$ and becomes larger with the wall-normal distance, i.e. at $y/D=0.5, 1.0$ and $2.0$.
The consistency of the trend between the estimation error and the standard deviation implies that the model provides `well-calibrated' uncertainty estimates, which is one of the strengths of SWAG compared to the other neural-network-based probabilistic approaches, as discussed by Maddox et al.~\cite{SWAGarxiv2019}.
{The same trend can also be observed from the distribution of standard deviation fields and the error fields depicted at the bottom of figure~\ref{fig:sqcy}.
Both fields show relatively high concentration on the region around vortical structures.}

\subsubsection{Example 2-2: Flow in a simplified urban environment}

\begin{figure}
  \centering
  \includegraphics[width=\textwidth]{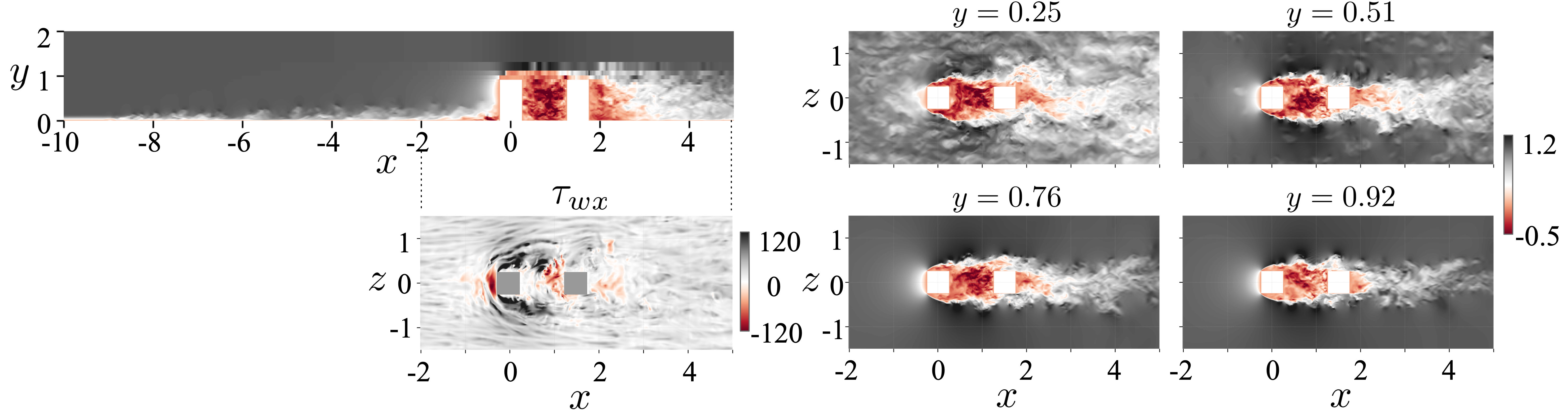}
  \caption{Representative snapshots of streamwise velocity in the urban-flow data.}
  \label{fig:urban_ex}
\end{figure}

Here, we examine the applicability of SWAG-based UQ to more practical and complex situations considering the turbulent flow through a simplified two-obstacle urban structure~\cite{urbanflow}.
The dataset was obtained from the well-resolved large-eddy simulation (LES) by Torres et al.~\cite{urbanflow}, who employed the spectral-element code Nek5000, which relies on a Gauss--Lobatto--Legendre (GLL) quadrature.
The friction Reynolds number upstream of the first obstacle is around $Re_\tau=280$ (based on $99\%$ boundary-layer thickness $\delta_{99}$ and friction velocity $u_{\tau}$), and around 103 million grid points are utilized for discretization.
A no-slip wall is considered at the bottom of the obstacles, i.e., $y=0$, while a prescribed freestream velocity (together with a homogeneous Neumann condition in $y$) is imposed at $y=2$.
The obstacles are located at $x=0$ and $1.5$, and their height is precisely the simulation length scale $h$.
The region around the obstacles, i.e., $-2<x<5$, is extracted as training data for a machine-learning model, as illustrated on the left side of figure~\ref{fig:urban_ex}.
The problem setting is similar to that of the square-cylinder wake example: the input is an $x-z$ plane of instantaneous streamwise wall-shear stress {$\tau_{wx}=\mu({\partial u}/{\partial y}|_{\rm wall})$} and the outputs are $x-z$ planes of streamwise velocity data at $y=\{0.25,0.51,0.76,0.92\}$.
We use 600 snapshots for training the model while 200 snapshots are utilized for the evaluation.

\begin{figure}[t]
  \centering
  \includegraphics[width=\textwidth]{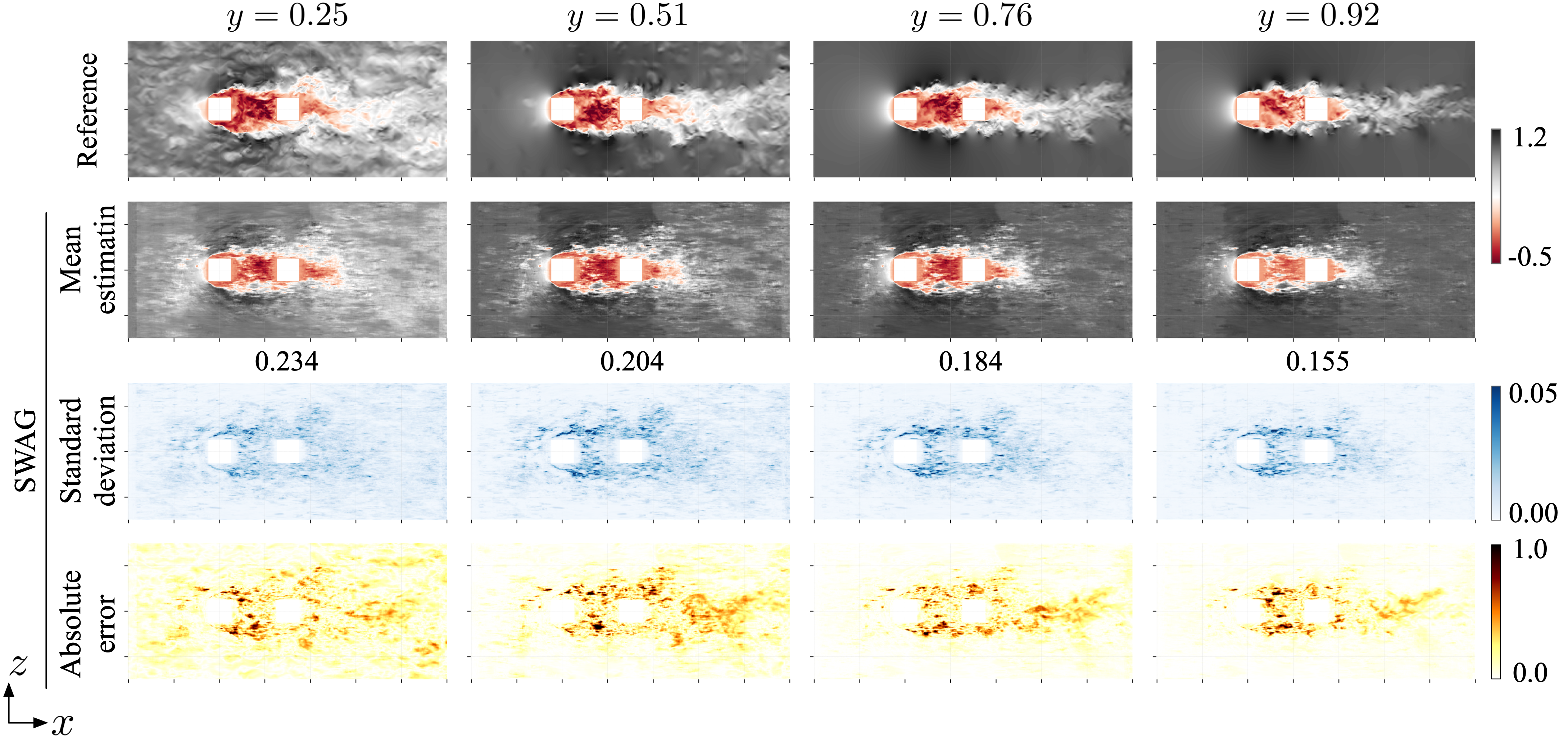}
  \caption{Estimated streamwise velocity fields from the streamwise wall-shear-stress input.
  {Fields of mean estimations, standard deviations, and absolute errors obtained through the present SWAG are depicted.}
  The values underneath the estimated field represent the $L_2$ error norm.
  }
  \label{fig:urban_swag}
\end{figure}

\begin{figure}
  \centering
  \includegraphics[width=0.6\textwidth]{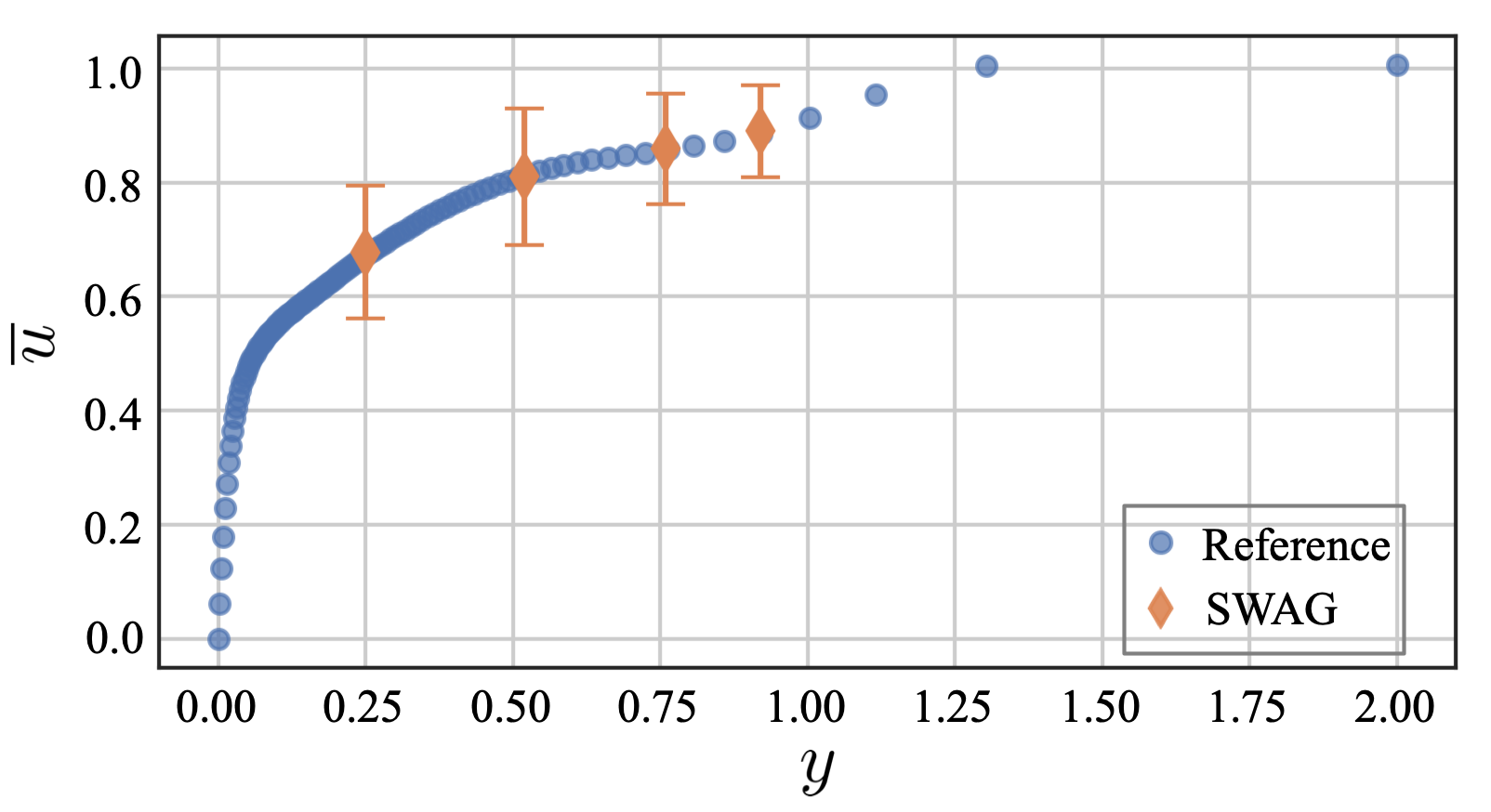}
  \caption{Relationship between the the mean velocity profile and the distance from the wall surface $y$.
  An errorbar for each plot of SWAG represents $20\sigma$ of the estimation at each section.}
  \label{fig:urms_y}
\end{figure}

The mean estimates and standard deviations are shown in figure~\ref{fig:urban_swag}.
The present CNN can estimate a far-field state with an $L_2$ error norm ranging from 0.15 to 0.24.
Contrary to the case with the square-cylinder wake, the estimation performance becomes better with the sections farther away from the input section.
This is because the complexity of the flow is reduced as the distance from the wall increases, with fewer large-scale features dominating the flow dynamics.
{The spatial distributions of the estimated standard deviation and the error present the similar trend with each other, as depicted at the bottom of figure~\ref{fig:urban_swag}.
Both fields have a high concentration in the region near the bodies which corresponds to a high velocity fluctuation.}

A relationship between the mean velocity profile $\overline{u}$ (temporally and spatially averaged at each $x-z$ cross section) and the distance from the wall is summarized in figure~\ref{fig:urms_y}.
The mean velocity profile obtained from the estimations of SWAG shows reasonable agreement with the reference data.
Moreover, the standard deviations of them are in a quite narrow range, i.e., under 1.0\% of the reference value, which implies that the SWAG-based estimation captures the spatio-temporal averaged value in a reasonable manner.

Furthermore, we also demonstrate the applicability of SWAG-based uncertainty quantification on low-dimensional variables for reduced-order modeling (ROM).
Since ROM, e.g., based on proper-orthogonal decomposition (POD)~\cite{Lumely1967} and autoencoders~\cite{HS2006,FHNMF2020}, play a crucial role to analyze, predict, and also control a high-dimensional system in low-dimensional spaces~\cite{taira_modalanalysis2017}, performing UQ on a latent space can also be regarded as an important extension of the method.
As an example, we take the SWAG-based sample outputs and obtain low-dimensional representations in terms of modal content.
We first perform POD for temporal snapshots of the mean estimation ${\bm y}_{\rm SWAG}$ of SWAG as follows:
\begin{equation}
    {\bm y}_{\rm SWAG}={\bm y}_{{\rm SWAG},0}+\sum_{i}a_{{\rm SWAG},i}\phi_{{\rm SWAG},i},
\end{equation}
where ${\bm y}_{{\rm SWAG},0}$ and $a_{{\rm SWAG},i}$ respectively represent a time-averaged field of the mean estimation and the temporal coefficients of the POD modes, respectively.
We then use this set of basis functions to obtain coefficients for the SWAG predictions for each weight sample:
\begin{equation}
    \tilde{\bm a}=\tilde{\bm y}^\prime{\bm \Phi}^{\rm T}_{\rm SWAG},
\end{equation}
where $\tilde{\bm a}\in{\mathbb R}^{n_{\rm mode}\times n_{\rm time}}$ is a temporal-coefficient matrix and ${\bm \Phi}_{\rm SWAG}$ $=[\phi_{{\rm SWAG},1},$ $\phi_{{\rm SWAG},2},$ $\cdots,$ $\phi_{{\rm SWAG},n_{\rm mode}}]$ is a matrix containing POD bases for each mode.
Note that $\tilde{\bm y}^\prime$ represents a fluctuation component of $\tilde{\bm y}$ which is associated with mean and standard deviation of the estimation, $\tilde{\bm y}={\bm y}_{\rm SWAG}+{\bm \sigma}_{\rm SWAG}$.

\begin{figure}
  \centering
  \includegraphics[width=\textwidth]{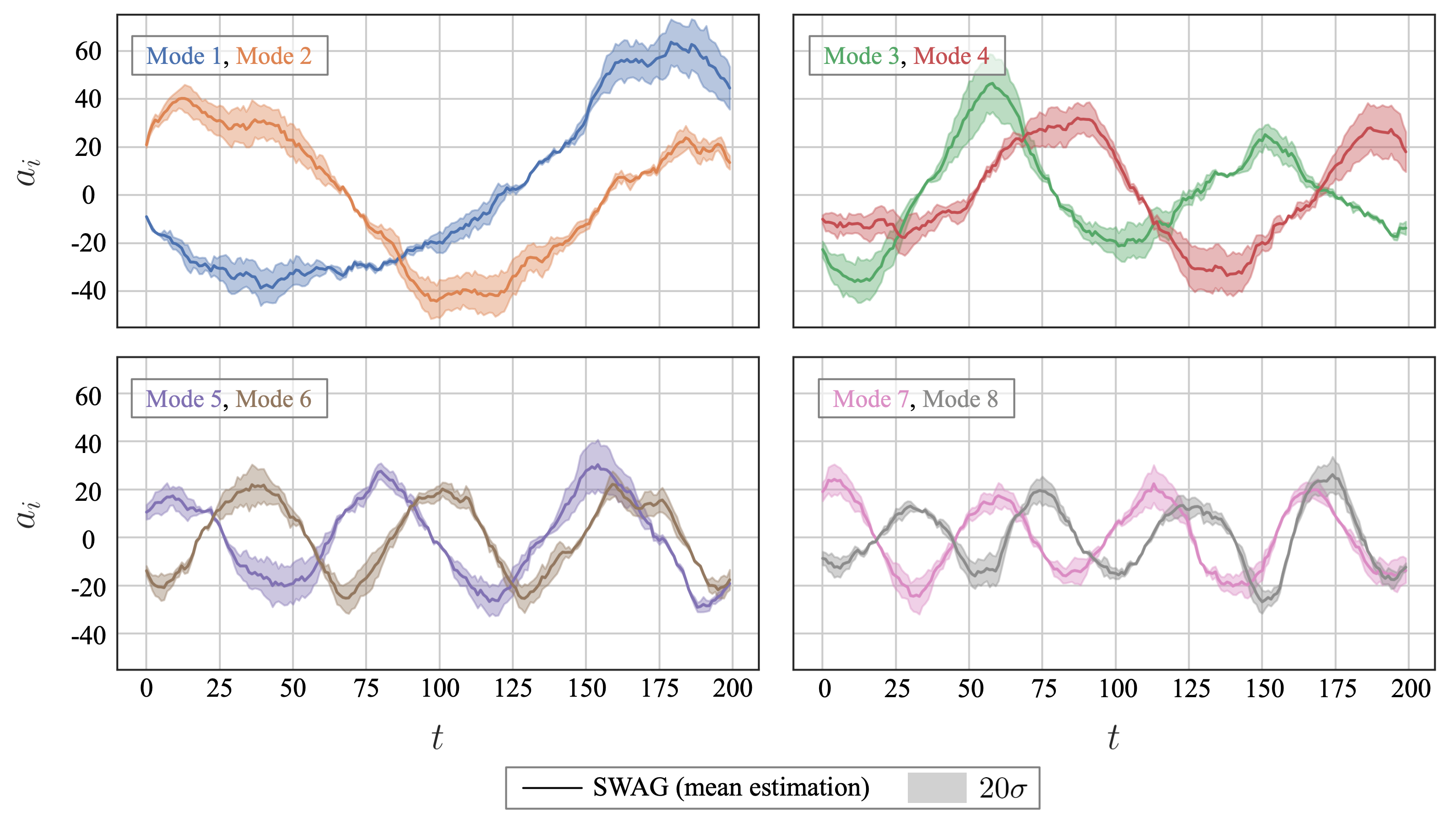}
  \caption{SWAG-based uncertainty quantification for POD coefficients of the estimated velocity field at $y=0.25$.}
  \label{fig:urban_alpha}
\end{figure}

Time series of the estimated temporal coefficients (of the SWAG-predicted standard deviation) for the first eight dominant POD modes are presented in figure~\ref{fig:urban_alpha}.
The uncertainty of the coefficients tends to become larger when the coefficient magnitudes increase.
This is likely because a distribution of velocity components in the snapshots with larger coefficients is distanced from that of snapshots at $a=0$, i.e., the mean flow, where a supervised-machine-learning model trained with $L_2$ minimization tends to perform better~\cite{FFT2021b}.
Although this demonstration is just an illustrative example, the uncertainty specific to modal content can also be assessed from the described method.

\section{Concluding remarks}
\label{sec:conclusion}

In this study, we investigated the applicability of Gaussian stochastic weight averaging (SWAG)-based uncertainty quantification (UQ) for neural-network-based fluid-flow approximation.
The present method particularly focuses on `epistemic' uncertainty, which characterizes the deficiencies in the training data in the context of deep neural networks.
We demonstrated SWAG-based UQ considering two types of machine-learning models: (i) MLP-CNN-based global-field estimator from sparse-sensor measurements and (ii) CNN-based far-field state estimator.
Through a global-field reconstruction task of cylinder wakes at various Reynolds numbers, we found that a distribution of weights can successfully be approximated from those collected during training with SGD.
In addition, epistemic uncertainty can be assessed via repeated sampling from this distribution.
A case in which the method fails to perform adequately, i.e. the DayMET dataset, was also investigated. 
Analyses for this case also demonstrated that the distribution of the weights and the smoothness of the loss surface can provide insight into the suitability of SWAG for a problem setting, and indeed, the quality of a neural-network architecture, which strongly relates to a multimodality of the loss surface of weights.
Investigations of the weight distributions reveal that deviations from a unimodal nature during the sampling phase led to insufficiently robust function approximation.
Although results for this experiment were improved with a customization of the optimization method, care should be taken depending on both problem and parameter settings.

A far-field estimation task was also considered for the present demonstration with flows around a square cylinder and simplified-urban-environment geometry.
We exemplified the characteristics of the well-calibrated SWAG, which showed a consistent trend for estimation capability and extracting confidence intervals~\cite{SWAGarxiv2019}.
SWAG samples can also be combined with a modal-decomposition technique to analyze fluctuations with respect to coherent structures. When coupled with a POD-postprocessing technique, variations in different basis could be compared. Subsequently, such analyses may be used for balancing approximation and compression capabilities using deep learning.
Future extensions of SWAG for deep learning in fluid dynamics will include utilizing uncertainty estimates for augmenting training data or for multifidelity function approximation and surrogate modeling.

\section*{Acknowledgments}

M.M., Ka.F, and Ko.F thank the support by JSPS KAKENHI Grant Number 18H03758 and 21H05007.
R.V. acknowledges the financial support by the G\"oran Gustafsson Foundation. 
This material is based upon work supported by the U.S. Department of Energy (DOE), Office of Science, Office of Advanced Scientific Computing Research, under Contract DE-AC02-06CH11357. This research was funded in part and used resources of the Argonne Leadership Computing Facility, which is a DOE Office of Science User Facility supported under Contract DE-AC02-06CH11357. 
This paper describes objective technical results and analysis. Any subjective views or opinions that might be expressed in the paper do not necessarily represent the views of the U.S. DOE or the United States Government.

\section*{Data availability}

The data that support the findings of this study are available from the corresponding author upon reasonable request.

\section*{Declaration of interest}

The authors report no conflict of interest.

\section*{Author contributions}

M.M., Ka.F, and R.M. designed research; M.M. and Ka.F performed research and analyzed data; R.V. and Ko.F. supervised. 
All authors reviewed the manuscript.


\bibliography{references}

\begin{thebibliography}{10}
\bibitem{kendall2017uncertainties}
A.~Kendall, Y.~Gal, What uncertainties do we need in {B}ayesian deep learning
  for computer vision?, Advances in neural information processing systems 30.
\end{thebibliography}

\end{document}